\documentclass[12pt,a4paper]{article}
\usepackage{a4wide}
\usepackage{bbm}                                                  
\usepackage{amsmath}
\usepackage{graphicx}

\newlength{\absize}
\newcommand{\nc}{\newcommand}
\nc{\hc}{\hbox {h.c.}} 
\def\lsim{\mathrel{\raise.3ex\hbox{$<$\kern-.75em\lower1ex\hbox{$\sim$}}}}
\def\gsim{\mathrel{\raise.3ex\hbox{$>$\kern-.75em\lower1ex\hbox{$\sim$}}}}

\def\gev{\;\hbox{GeV}}
\def\tev{\;\hbox{TeV}}

\def\cb{c_\beta}
\def\sb{s_\beta}

\renewcommand{\Re}{\mbox{Re\thinspace}}
\renewcommand{\Im}{\mbox{Im\thinspace}}

\newcommand{\half}{{\textstyle\frac{1}{2}}}
\newcommand{\fourth}{{\textstyle\frac{1}{4}}}
\allowdisplaybreaks %!!!!

\def\i11{{\mathbbm 1}}

\begin{document}

\thispagestyle{empty}
\renewcommand{\thefootnote}{\fnsymbol{footnote}}
\newpage\normalsize
\pagestyle{plain}
\setlength{\baselineskip}{4ex}\par
\setcounter{footnote}{0}
\renewcommand{\thefootnote}{\arabic{footnote}}
\newcommand{\preprint}[1]{%
\begin{flushright}
\setlength{\baselineskip}{3ex} #1
\end{flushright}}
\renewcommand{\title}[1]{%
\begin{center}
\LARGE #1
\end{center}\par}
\renewcommand{\author}[1]{%
\vspace{2ex}
{\Large
\begin{center}
 \setlength{\baselineskip}{3ex} #1 \par
\end{center}}}
\renewcommand{\thanks}[1]{\footnote{#1}}
\renewcommand{\abstract}[1]{%
\vspace{2ex}
\normalsize
\begin{center}
\centerline{\bf Abstract}\par
\vspace{2ex}
\parbox{\absize}{#1\setlength{\baselineskip}{2.5ex}\par}
\end{center}}

\begin{flushright}
NORDITA-2009-24\\
IFT-09-05
\end{flushright}

\vspace*{4mm} 

\title{Natural Multi-Higgs Model \\ 
with Dark Matter and CP Violation} \vfill

\author{B. Grzadkowski,$^{a,}$\footnote{E-mail: Bohdan.Grzadkowski@fuw.edu.pl}
O. M. Ogreid$^{b,}$\footnote{E-mail: omo@hib.no}
P. Osland,$^{c,}$\footnote{E-mail: Per.Osland@ift.uib.no} }

\begin{center}
$^{a}$Institute of Theoretical Physics, University of Warsaw, \\ Ho\.za 69, 
PL-00-681 Warsaw, Poland\\
$^{b}$Bergen University College, Bergen, Norway \\
$^{c}$Department of Physics and Technology, University of Bergen, \\
Postboks 7803, N-5020  Bergen, Norway
\end{center}
\vfill
%%%%%%%%%%%%%%%%%%%%%%%%%%%%%%%%%%%%%%%%%%%%%%%%
\abstract{We explore an extension of the Inert Doublet Model which allows
also for CP violation in the Higgs sector. This necessitates two non-inert
doublets. The lightest neutral scalar of the inert doublet is a candidate for
dark matter. Scanning over parameters we preserve the abundance of the dark
matter in agreement with the WMAP data.  We also impose all relevant collider
and theoretical constraints to determine the allowed parameter space for which
both the dark matter is appropriate and CP is violated. In addition we find
regions where the cut-off of the model originating from naturality arguments
can be substantially lifted compared to its Standard Model value, reaching
$\sim 2-3\tev$. }

\vspace*{20mm} \setcounter{footnote}{0} \vfill

\newpage
\setcounter{footnote}{0}
\renewcommand{\thefootnote}{\arabic{footnote}}

%%%%%%%%%%%%%%%%%%%%%%%%%%%%%%%%%%%%%%%%%%%%%%%%%%%%%%%%%%%%%%%%%%%%%%%%%%%%%
\section{Introduction}
%%%%%%%%%%%%%%%%%%%%%%%%%%%%%%%%%%%%%%%%%%%%%%%%%%%%%%%%%%%%%%%%%%%%%%%%%%%%%

It is widely believed that the standard model (SM) of electroweak interactions
is only an effective low-energy theory valid below a certain energy scale
$\Lambda$, which is supposed to be of the order of 1~TeV.  This view is based
on the fact that radiative corrections, $\delta m_h^2$, to the Higgs boson
mass squared ($m_h^2$), tend to increase the mass up to $\Lambda$, implying a
necessary fine tuning.  This is the celebrated little hierarchy problem or the
``LEP paradox" \cite{Barbieri:2000gf}. In order to retain a meaningful
perturbative expansion above $\sim 1\tev$, a high level of fine tuning between
$m_h$ and $\Lambda$ is necessary to suppress $\delta m_h^2$ relative to
$m_h^2$, see e.g.\ \cite{Kolda:2000wi}.  Other well know problems of the SM
are the lack of a candidate for the dark matter (DM) and too little CP
violation (CPV) that could make the electroweak baryogenesis
viable~\cite{Bernreuther:2002uj}. In that context the SM scalar sector has to
be modifed as the phase transition within a single Higgs doublet is too slow
for baryogenesis~\cite{Bernreuther:2002uj}.

Our goal here is to outline a model that ameliorates the
little hierarchy problem (lifting $\Lambda$ at least to $\sim 2\tev$) 
while providing extra sources of CP violation
needed for baryogenesis as well as a realistic abundance of dark matter. 
We will focus on extending 
the Higgs sector of the SM by adding extra Higgs doublets 
since that could also help to make the electroweak
phase transition fast enough, see \cite{Fromme:2006cm}. 

In general there are two possibilities to alleviate the little hierarchy
problem: (i) suppression of the radiative corrections, and/or (ii) an increase
of the Higgs mass. Well known examples of the first strategy are
supersymmetric extensions of the SM, however in fact that could be achieved to
some extent even through very modest means, e.g.\ by introducing just one
extra real scalar singlet to the SM~\cite{Grzadkowski:2009mj} (although some
fine tuning is necessary).  The second possibility has recently been followed
by Barbieri, Hall and Rychkov \cite{Barbieri:2006dq}\footnote{The model was
proposed earlier in \cite{Deshpande:1977rw} as a possible solution to the DM
problem, its collider phenomenology was then discussed in
\cite{Cao:2007rm}.}. The idea is to introduce an extra scalar doublet $\eta$
(the inert doublet) which does not couple to the SM fermions as a consequence
of a $Z_2$ symmetry: $\eta \to -\eta$ (all other fields are neutral under the
symmetry). Since it is required that $\langle \eta \rangle = 0$, the symmetry
remains unbroken, therefore it can provide a DM candidate. Since the inert
doublet contributes to gauge-boson two-point Green's functions the SM-like
Higgs boson could be as heavy as $400-600\gev$, ameliorating the little
hierarchy problem this way.  Also the DM constraints could be satisfied
choosing masses of the scalar and pseudoscalar components of the inert Higgs
of the order of $80\gev$. The analysis of \cite{LopezHonorez:2006gr} reveals
also another solution for the DM candidate, such that scalar and pseudoscalar
masses are much heavier, $\gsim 500\gev$.  The model in very simple terms
avoids the little hierarchy problem for the lighter (inert) scalars as they
just do not couple to fermions, so in particular not to the top quark. The
only drawback is that the model is so restricted by the $Z_2$ symmetry that it
does not allow for CP violation in the Higgs potential, this is the issue that
we would like to address here. There are two simple extensions of the Inert
Doublet Model (IDM) that can accommodate CP violation:
\begin{itemize}
\item Combining the standard 2HDM with an inert scalar
doublet $\eta$ (in other words replacing the SM Higgs doublet of the IDM by 
{\it two} doublets). 
\item Adding a complex singlet scalar to the IDM.
\end{itemize}

This paper is devoted to the first of these extensions. It is worth emphasizing
that this scenario is not just a simple sum of the 2HDM and the IDM. Although 
some theoretical and experimental constraints which are applicable here,
are (to leading order) identical to those of the 2HDM (just because the inert doublet
does not couple to fermions), there are also important exceptions. These
concern the oblique parameters $T$ and $S$, the amount of DM and
the issue of positivity of the scalar potential (vacuum stability):
\begin{itemize}
\item
The extra inert degrees of freedom transform as an SU(2) doublet, so they couple to 
the vector bosons and therefore contribute to the oblique parameters, modifying 
the standard 2HDM predictions.
\item
The neutral components of the inert doublet are candidates for DM, and since
the inert doublet couples to the 2HDM doublets, the amplitudes for
DM annihilation are in general influenced in a non-trivial way by the extension of 
the non-inert sector. 
\item
Even under simplifying assumptions, the scalar potential for the 2HDM extended by the
inert doublet has a rich structure, so that the condition for positivity 
is much more involved than a simple superposition of conditions needed for the 2HDM and
the IDM separately.
\end{itemize}

The paper is organized as
follows. In Sec.~\ref{Sec:def-model} we introduce the model and define some
notation. In Sec.~\ref{Sec:benchmarks} we define some benchmarks for the inert
sector, and in Sec.~\ref{Sec:strategy} we present the strategy adopted to
search for allowed regions in the parameter space of the model.
Secs.~\ref{Sec:theory-constraints} and \ref{Sec:exp-constraints} are devoted
to reviews of theoretical and experimental constraints. In
Sec.~\ref{Sec:results} we show some regions of parameters of the model that
are compatible with all the constraints, and in Sec.~\ref{Sec:summary} we
summarize.

Technical details on positivity, CP conservation and necessary basis
transformations are collected in appendices A, B and C.

%%%%%%%%%%%%%%%%%%%%%%%%%%%%%%%%%%%%%%%%%%%%%%%%%%%%%%%%%%%%%%%%%%%%%%%%%%%%%
\section{Inert-plus-two-doublet model: IDM2}
\label{Sec:def-model}
\setcounter{equation}{0}
%%%%%%%%%%%%%%%%%%%%%%%%%%%%%%%%%%%%%%%%%%%%%%%%%%%%%%%%%%%%%%%%%%%%%%%%%%%%%
%%%%%%%%%%%%%%%%%%%%%%%%%%%%%%%%%%%%%%%%%%%%%%%%%%%%%%%%%%%%%%%%%%%%%%%%%%%%%
\subsection{The potential}
%%%%%%%%%%%%%%%%%%%%%%%%%%%%%%%%%%%%%%%%%%%%%%%%%%%%%%%%%%%%%%%%%%%%%%%%%%%%%

Introduction of two doublets, $\Phi_{1,2}$ leads in general to Flavor-Changing
Neutral Currents in Yukawa couplings. To avoid those one can impose an extra
$Z'_2$ symmetry such that $\Phi_1 \to -\Phi_1$ and $u_R\to -u_R$ (all other
fields are neutral). The model then has $Z_2\times Z'_2$, where the first
factor is the inert-doublet $Z_2$: $\eta \to -\eta$ (all other fields are
neutral). The potential reads
\begin{equation} \label{Eq:fullpot}
V(\Phi_1,\Phi_2,\eta) 
= V_{12}(\Phi_1,\Phi_2) + V_3(\eta) + V_{123}(\Phi_1,\Phi_2,\eta)
\end{equation}
where
\begin{align} 
V_{12}(\Phi_1,\Phi_2) &= -\frac12\left\{m_{11}^2\Phi_1^\dagger\Phi_1 
+ m_{22}^2\Phi_2^\dagger\Phi_2 + \left[m_{12}^2 \Phi_1^\dagger \Phi_2 
+ \hc\right]\right\} \nonumber \\
& + \frac{\lambda_1}{2}(\Phi_1^\dagger\Phi_1)^2 
+ \frac{\lambda_2}{2}(\Phi_2^\dagger\Phi_2)^2
+ \lambda_3(\Phi_1^\dagger\Phi_1)(\Phi_2^\dagger\Phi_2) 
+ \lambda_4(\Phi_1^\dagger\Phi_2)(\Phi_2^\dagger\Phi_1) 
\nonumber \\
& + \frac12\left[\lambda_5(\Phi_1^\dagger\Phi_2)^2 + \hc\right],
\label{v12} \\
V_3(\eta) &= m_\eta^2\eta^\dagger \eta + \frac{\lambda_\eta}{2} 
(\eta^\dagger \eta)^2,
\label{v3} \\
V_{123}(\Phi_1,\Phi_2,\eta) 
&=
\lambda_{1133} (\Phi_1^\dagger\Phi_1)(\eta^\dagger \eta)
+\lambda_{2233} (\Phi_2^\dagger\Phi_2)(\eta^\dagger \eta) \nonumber  \\
& +\lambda_{1331}(\Phi_1^\dagger\eta)(\eta^\dagger\Phi_1) 
+\lambda_{2332}(\Phi_2^\dagger\eta)(\eta^\dagger\Phi_2) \nonumber  \\
& 
+\half\left[\lambda_{1313}(\Phi_1^\dagger\eta)^2 +\hc \right]  
+\half\left[\lambda_{2323}(\Phi_2^\dagger\eta)^2 +\hc \right].
\label{v123}
\end{align}
Here, $\lambda_{1133}$, $\lambda_{2233}$, $\lambda_{1331}$ and
$\lambda_{2332}$ are real, whereas $\lambda_{1313}$ and $\lambda_{2323}$ can
be complex.  Disregarding $\Phi_2$, the correspondence with the notation of
\cite{Barbieri:2006dq} would be
\begin{equation}
(m_\eta,\lambda_\eta,\lambda_{1133},\lambda_{1331},\lambda_{1313})
\leftrightarrow
(\mu_2,2\lambda_2,\lambda_3,\lambda_4,\lambda_5).
\end{equation}

In $V_1(\Phi_1,\Phi_2)$ we have allowed for soft breaking of $Z'_2$ in order
to preserve the chance of CP violation in the potential while we do not allow
for any breaking of $Z_2$ in order to have a stable lightest component of
$\eta$ as a DM candidate. Note that, as a consequence of the unbroken $Z_2$,
there is no mixing in mass terms between $\Phi_{1,2}$ and $\eta$. It is worth
realizing that, since $\eta$ does not couple to quarks, there are no
constraints on the charged inert Higgs mass from the $b\to s \gamma$ decay.

%%%%%%%%%%%%%%%%%%%%%%%%%%%%%%%%%%%%%%%%%%%%%%%%%%%%%%%%%%%%%%%%%%%%%%%%%%%%%
\subsection{2HDM mass eigenstates}  
%%%%%%%%%%%%%%%%%%%%%%%%%%%%%%%%%%%%%%%%%%%%%%%%%%%%%%%%%%%%%%%%%%%%%%%%%%%%%
In the (non-inert) 2HDM sector of the model, we denote the doublets
(in a basis where both have a v.e.v.)
\begin{equation} 
\Phi_1=\left(
\begin{array}{c}\varphi_1^+\\ (v_1+\eta_1+i\chi_1)/\sqrt{2}
\end{array}\right), \quad
\Phi_2=\left(
\begin{array}{c}
\varphi_2^+\\ (v_2+\eta_2+i\chi_2)/\sqrt{2}
\end{array}
\right),
\label{Obasis}
\end{equation}
and adopt the mixing matrix $R$, defined by
\begin{equation} \label{Eq:R-def}
\begin{pmatrix}
H_1 \\ H_2 \\ H_3
\end{pmatrix}
=R
\begin{pmatrix}
\eta_1 \\ \eta_2 \\ \eta_3
\end{pmatrix},
\end{equation}
satisfying
\begin{equation}
\label{Eq:cal-M}
R{\cal M}^2R^{\rm T}={\cal M}^2_{\rm diag}={\rm diag}(M_1^2,M_2^2,M_3^2),
\end{equation}
and parametrized in terms of three rotation angles $\alpha_i$ as
\begin{equation}     \label{Eq:R-angles}
R
=\begin{pmatrix}
c_1\,c_2 & s_1\,c_2 & s_2 \\
- (c_1\,s_2\,s_3 + s_1\,c_3) 
& c_1\,c_3 - s_1\,s_2\,s_3 & c_2\,s_3 \\
- c_1\,s_2\,c_3 + s_1\,s_3 
& - (c_1\,s_3 + s_1\,s_2\,c_3) & c_2\,c_3
\end{pmatrix}
\end{equation}
with $c_i=\cos\alpha_i$, $s_i=\sin\alpha_i$.
In Eq.~(\ref{Eq:R-def}), $\eta_3 \equiv -\sin\beta\chi_1+\cos\beta\chi_2$
is the combination of $\chi_i$'s which is 
orthogonal to the neutral Nambu--Goldstone boson.
Here, $\tan\beta\equiv v_2/v_1$.

We also define a mass parameter
$\mu^2\equiv (v^2/2 v_1 v_2) \Re m_{12}^2$, and note
the following useful relation:
\begin{equation}
\Im m_{12}^2=\Im\lambda_5v_1v_2.
\label{m12}
\end{equation}

%%%%%%%%%%%%%%%%%%%%%%%%%%%%%%%%%%%%%%%%%%%%%%%%%%%%%%%%%%%%%%%%%%%%%%%%%%%%%
\subsection{Inert-sector mass eigenstates}
%%%%%%%%%%%%%%%%%%%%%%%%%%%%%%%%%%%%%%%%%%%%%%%%%%%%%%%%%%%%%%%%%%%%%%%%%%%%%
Components of the inert doublet are defined as follows
\begin{equation}
\eta = \left(
\begin{array}{c}
 \eta^+ \\ (S+iA)/\sqrt{2} 
\end{array}
\right).
\label{eta}
\end{equation}
The masses of the inert scalars will be given by expressions analogous to
those of \cite{Barbieri:2006dq, LopezHonorez:2006gr}:
\begin{align}
M^2_{\eta^\pm}
&=m_\eta^2+\half\Delta_\text{ch}^2, \nonumber \\
M^2_S
&=m_\eta^2+\half[\Delta_\text{ch}^2+\Delta_0^2+\Delta_\text{split}^2], 
\nonumber \\
M^2_A
&=m_\eta^2+\half[\Delta_\text{ch}^2+\Delta_0^2-\Delta_\text{split}^2], 
\end{align}
where we have introduced the abbreviations
\begin{align}
\Delta_\text{ch}^2
&=\lambda_{1133}\,v_1^2+\lambda_{2233}\,v_2^2, \nonumber \\
\Delta_0^2
&=\lambda_{1331}\,v_1^2+\lambda_{2332}\,v_2^2, \nonumber \\
\Delta_\text{split}^2
&=\Re\lambda_{1313}\,v_1^2+\Re\lambda_{2323}\,v_2^2
\end{align}

Adopting the simplifying assumptions (denoted ``dark democracy'')
\begin{align} \label{Eq:DarkDemocracy}
\lambda_a\equiv \lambda_{1133}&=\lambda_{2233}, \nonumber \\
\lambda_b\equiv \lambda_{1331}&=\lambda_{2332}, \nonumber \\ 
\lambda_c\equiv \lambda_{1313}&=\lambda_{2323} \text{ (real)},
\end{align}
the masses can be written as:
\begin{align}
M^2_{\eta^\pm}
&=m_\eta^2+\half\lambda_a\,v^2, \nonumber \\
M^2_S
&=m_\eta^2+\half(\lambda_a+\lambda_b+\lambda_c)v^2
=M^2_{\eta^\pm}+\half(\lambda_b+\lambda_c)v^2, \nonumber \\
M^2_A
&=m_\eta^2+\half(\lambda_a+\lambda_b-\lambda_c)v^2
=M^2_{\eta^\pm}+\half(\lambda_b-\lambda_c)v^2.
\label{inmass}
\end{align}
As a consequence of the assumptions (\ref{Eq:DarkDemocracy}), there are
no trilinear couplings $H^\pm\eta^\mp S$ or $H^\pm\eta^\mp A$.
%%%%%%%%%%%%%%%%%%%%%%%%%%%%%%%%%%%%%%%%%%%%%%%%%%%%%%%%%%%%%%%%%%%%%%%%%%%%%
\subsection{Stability of the potential}
%%%%%%%%%%%%%%%%%%%%%%%%%%%%%%%%%%%%%%%%%%%%%%%%%%%%%%%%%%%%%%%%%%%%%%%%%%%%%

The condition for positivity of $V$ is discussed in Appendix~A for the general
potential, Eq.~(\ref{Eq:fullpot}).  In our numerical applications we will
limit ourselves to the case of ``dark democracy" defined in
(\ref{Eq:DarkDemocracy}).  We find that for this special case,
Eqs.~(\ref{constraints}) and (\ref{Eq:positivity-cond.2}) must be satisfied
for positivity. However, we restrict ourselves even further by requiring
$V_{12}$, $V_{3}$ and $V_{123}$ separately to be positive. Then, in addition
to the familiar constraint on
$V_{12}$~\cite{Deshpande:1977rw,Nie:1998yn,Kanemura:1999xf} and $V_{3}$, we
obtain the following condition:
\begin{equation}
\lambda_a\geq \max (0,- 2\lambda_b, - \lambda_b \pm \lambda_c),
\label{posit}
\end{equation}
implying $m_\eta<M_{\eta^\pm}$.
This amounts to a strong constraint on the splitting of the inert-sector spectrum, not present
in the full treatment of positivity.

The input parameters in the inert sector are defined by specifying scalar
masses $(M_S,M_A,M_{\eta^\pm})$ together with $m_\eta$, so that the quartic
couplings $\lambda_a, \lambda_b$ and $\lambda_c$ can be determined via
(\ref{inmass}).  Here we will consider cases (the profile 3 is the only
exception, see Sec.~\ref{Subsec:_Light_DM_particle}), with masses ordered as
follows: $M_S<M_A<M_{\eta^\pm}$. This is motivated by the fact that a positive
contribution to the electroweak precision observable $T$
from the inert sector (see Sec.~\ref{Sec:exp-constraints})
makes it easier to allow for heavy 2HDM masses. For this case it is easy to
show that the ordering and the positivity condition (\ref{posit}) leave a
certain non-empty allowed region in the space of
$(\lambda_a,\lambda_b,\lambda_c)$, namely $\lambda_a > 0$, $\lambda_b < 0$,
$\lambda_a > 2 |\lambda_b|$ together with $\lambda_c<0$ and $|\lambda_b| >
|\lambda_c|$.  Then the requirement of having the right amount of dark matter
(see next section) imposes additional constraints on the masses (or
equivalently on the quartic couplings), resulting in a relatively small region
of allowed $(\lambda_a,\lambda_b,\lambda_c)$.  If $M_S<M_{\eta^\pm}<M_A$, then
the constraints are less tight.

%%%%%%%%%%%%%%%%%%%%%%%%%%%%%%%%%%%%%%%%%%%%%%%%%%%%%%%%%%%%%%%%%%%%%%%%%%%%%
\section{Benchmarks}
\label{Sec:benchmarks}
\setcounter{equation}{0}
%%%%%%%%%%%%%%%%%%%%%%%%%%%%%%%%%%%%%%%%%%%%%%%%%%%%%%%%%%%%%%%%%%%%%%%%%%%%%
In numerical studies we will assume that $S$ is the lightest neutral scalar,
$M_S<M_A$. The first profiles have the DM candidate around 75~GeV, a favoured
value \cite{Barbieri:2006dq,LopezHonorez:2006gr}.  For the heavier neutral
partner, we consider a few options at, or slightly above $M_A=110~\text{GeV}$,
which is the lower limit compatible with LEP2 data
\cite{Lundstrom:2008ai}. 

In a recent study by Lopez Honorez et al.\ \cite{LopezHonorez:2006gr}, the
splitting amongst inert-sector scalar masses was kept fixed, while a scan over
the parameter $\mu_2$ (corresponding to our $m_\eta$) and the DM particle mass
($M_S$ in our notation) was performed. 
In the region of heavy DM particles, a detailed study has also 
been performed in \cite{Hambye:2009pw}.

We shall instead consider two discrete
sets of ``dark'' profiles, to be specified below.
In the first set of ``dark profiles'', we keep the DM particle light.  For
trial values of the masses and $m_\eta$ (inspired by the results of
\cite{LopezHonorez:2006gr}), we estimate the amount of dark matter from
micrOMEGAs2.2 \cite{Belanger:2006is,Belanger:2008sj}, in the IDM version
developed by Lopez Honorez et al.\ \cite{LopezHonorez:2006gr}.  That version
has only {\it one} $Z_2$-even doublet, whereas we here consider two such
doublets, $\Phi_1$ and $\Phi_2$, (with many more ``free'' parameters).
Therefore the calculation of amplitudes for various DM annihilation channels
is more complicated. 

The mass of the charged partner, $M_{\eta^\pm}$, and the inert-sector mass
parameter $m_\eta$ have been chosen such that a reasonable amount of dark
matter is obtained for at least one set of the parameters
(\ref{Eq:non-inert-masses}).  In the context of dark matter, the parameter
$m_\eta$ is important, since a particular choice for the inert neutral scalar
masses $M_S$ and $M_A$ together with $m_\eta$ constrain
$(\lambda_a,\lambda_b,\lambda_c)$ which, in turn, are responsible for the
annihilation of dark matter into the visible sector, see
\cite{LopezHonorez:2006gr}. In the original IDM, the essential parameter
determining the trilinear coupling among two DM-particles and the SM Higgs
boson, is
\begin{equation} \label{Eq:lambda_L}
\lambda_L\equiv \half(\lambda_a+\lambda_b+\lambda_c).
\end{equation}

In the present model, the trilinear coupling between two $S$-particles and a
neutral Higgs boson is determined by
\begin{equation}
\lambda_L(v_1\eta_1+v_2\eta_2)SS.
\end{equation}
Projecting out the coupling to a particular neutral Higgs boson, we find:
\begin{equation}
SSH_j: \quad
F_{SSj}\lambda_L, \quad
\text{with} \quad
F_{SSj}=\cos\beta R_{j1}+\sin\beta R_{j2},
\end{equation}
where the pre-factor satisfies $|F_{SSj}|\leq1$, since $R$ is unitary. In
particular, $F_{SS1}=\cos(\beta-\alpha_1)\cos\alpha_2$.

Similarly, the four-point coupling involving two $S$-particles and two neutral
Higgs bosons is determined by
\begin{equation}
\fourth[(\lambda_a+\lambda_b+\lambda_c)(\eta_1^2+\eta_2^2)
+(\lambda_a+\lambda_b-\lambda_c)(\chi_1^2+\chi_2^2)]SS.
\end{equation}
Projecting out the coupling to the lightest Higgs boson, we find:
\begin{equation}
SSH_1H_1:\quad
\half(\lambda_L-\lambda_c R_{13}^2)
=\half(\lambda_L-\lambda_c\sin^2\alpha_2).
\end{equation}
In our estimates of the DM density, we shall follow the approach of
\cite{LopezHonorez:2006gr}, taking $\lambda_L$ as the relevant parameter.
The DM values presented in this work are based on the full tree-level
calculation performed using micrOMEGAs2.2
with all intermediate and final states originating from the rich 2HDM
structure of the model.\footnote{Within  the approximate treatment
of positivity, and keeping only the lowest mass state of the 2HDM,
the full result can be mimicked by an appropriate tuning of $m_\eta$,
which in turn amounts to a tuning of $\lambda_a$.}

%%%%%%%%%%%%%%%%%%%%%%%%%%%%%%%%%%%%%%%%%%%%%%%%%%%%%%%%%%%%%%%%%%%%%%%%%%%%%
\subsection{2HDM masses}
\label{Subsec:_2HDM_masses}
%%%%%%%%%%%%%%%%%%%%%%%%%%%%%%%%%%%%%%%%%%%%%%%%%%%%%%%%%%%%%%%%%%%%%%%%%%%%%

For each of the dark profiles, we consider the following mass parameters of
the (non-inert) 2HDM sector:
\begin{subequations} \label{Eq:non-inert-masses}
\begin{alignat}{4}
&{\rm Set~A}:&\quad
&(M_1,M_2)=(100,300)~\text{GeV}, &\quad &\mu=200~\text{GeV},
\label{setA}\\
&{\rm Set~B}:&\quad
&(M_1,M_2)=(200,400)~\text{GeV}, &\quad &\mu=400~\text{GeV},
\label{setB}\\
&{\rm Set~C}:&\quad
&(M_1,M_2)=(400,500)~\text{GeV}, &\quad &\mu=400~\text{GeV}.
\label{setC}
\end{alignat}
\end{subequations}
A non-zero value for $\mu$ is adopted, in order to accommodate the unitarity
constraints limiting quartic couplings. We avoid degeneracy of $M_1$ and
$M_2$, since that would be a source of potential difficulties to produce CP
violation, see Sec.~\ref{subsec:CP-Violation} for a detailed discussion.

%%%%%%%%%%%%%%%%%%%%%%%%%%%%%%%%%%%%%%%%%%%%%%%%%%%%%%%%%%%%%%%%%%%%%%%%%%%%%
\subsection{Light DM particle}
\label{Subsec:_Light_DM_particle}
%%%%%%%%%%%%%%%%%%%%%%%%%%%%%%%%%%%%%%%%%%%%%%%%%%%%%%%%%%%%%%%%%%%%%%%%%%%%%
We shall consider the following ``light'' DM profiles:
\begin{alignat}{4} \label{Eq:light-profiles}
&\text{Profile 1}: &\quad 
&M_S=75~\text{GeV}, &\quad 
&M_A=110~\text{GeV},&\quad 
&M_{\eta^\pm}=112~\text{GeV}, \nonumber \\
&\text{Profile 1'}: &\quad
&M_S=77~\text{GeV}, &\quad
&M_A=110~\text{GeV}, &\quad
&M_{\eta^\pm}=112~\text{GeV}, \nonumber \\
&\text{Profile 2}: &\quad
&M_S=75~\text{GeV}, &\quad
&M_A=120~\text{GeV}, &\quad
&M_{\eta^\pm}=125~\text{GeV}, \nonumber \\
&\text{Profile 3}: &\quad
&M_S=75~\text{GeV}, &\quad
&M_A=120~\text{GeV}, &\quad
&M_{\eta^\pm}=85~\text{GeV}, \nonumber \\
&\text{Profile 4}: &\quad
&M_S=100~\text{GeV}, &\quad
&M_A=110~\text{GeV}, &\quad
&M_{\eta^\pm}=115~\text{GeV}, \nonumber \\
&\text{Profile 5}: &\quad
&M_S=120~\text{GeV}, &\quad
&M_A=125~\text{GeV}, &\quad
&M_{\eta^\pm}=130~\text{GeV}.
\end{alignat}

%%%%%%%%%%%%%%%%%%%%%%%%%%%%%%%%%%%%%%%%%%%%%%%%%%%%%%%%%%%%%%%%%%%%%%%%%%%%%
\subsection{Heavier DM particle}
%%%%%%%%%%%%%%%%%%%%%%%%%%%%%%%%%%%%%%%%%%%%%%%%%%%%%%%%%%%%%%%%%%%%%%%%%%%%%
We also consider some profiles where the two neutral inert-particle masses are
higher, and rather close, another domain favoured by
\cite{LopezHonorez:2006gr}:
\begin{alignat}{4}
&\text{Profile 11}: &\quad 
&M_S=500~\text{GeV}, &\quad 
&M_A=501~\text{GeV},&\quad 
&M_{\eta^\pm}=502~\text{GeV}, \nonumber \\
&\text{Profile 12}: &\quad 
&M_S=600~\text{GeV}, &\quad 
&M_A=601~\text{GeV},&\quad 
&M_{\eta^\pm}=602~\text{GeV}, \nonumber \\
&\text{Profile 13}: &\quad 
&M_S=800~\text{GeV}, &\quad 
&M_A=802~\text{GeV},&\quad 
&M_{\eta^\pm}=804~\text{GeV}, \nonumber \\
&\text{Profile 14}: &\quad 
&M_S=1000~\text{GeV}, &\quad 
&M_A=1002~\text{GeV},&\quad 
&M_{\eta^\pm}=1005~\text{GeV}.
\end{alignat}

The latter profiles have a high degree of degeneracy among the masses. This is
required in order to have the correct amount of dark matter, but will also
minimize the contribution to the electroweak observable $T$.
For a detailed study of the IDM in the high-mass regime, see \cite{Hambye:2009pw}.
%%%%%%%%%%%%%%%%%%%%%%%%%%%%%%%%%%%%%%%%%%%%%%%%%%%%%%%%%%%%%%%%%%%%%%%%%%%%%
\section{Search Strategy}
\label{Sec:strategy}
\setcounter{equation}{0}
%%%%%%%%%%%%%%%%%%%%%%%%%%%%%%%%%%%%%%%%%%%%%%%%%%%%%%%%%%%%%%%%%%%%%%%%%%%%%

In general, our goal will be to verify that within the IDM2 model one can 
accommodate the following features 
\begin{itemize}
\item a ``heavy'' lightest neutral scalar (so that the naturalness problem is 
alleviated),
\item at least one neutral scalar odd under $Z_2$ consistent with the
present limits on the DM abundance,
\item CP violation in the potential (introduced via $m_{12}^2$ and
$\lambda_5$).
\end{itemize}

First we choose the following input parameters for the 2HDM: $\tan
\beta$, $M_1$, $M_2$, $M_{H^\pm}$ and $\mu$, together with $(\alpha_1,
\alpha_2,\alpha_3)$.  All remaining parameters of the 2HDM
sector are calculable in terms of those chosen above \cite{Khater:2003wq} (see
also \cite{ElKaffas:2007rq}).  For the inert sector we choose $M_S\sim
70-80\gev$ or $\gsim 500\gev$ (DM candidate), $M_A$, $M_{\eta^\pm}$ and
$m_{\eta}$ (needed for determination of relevant quartic couplings in
$V_{123}$ which are necessary for calculating the DM abundance).

Then the following strategy will be applied while determining allowed regions
in the parameter space of the model:
\begin{itemize}
\item 
We fix $\tan \beta$, $M_{H^\pm}$, sets of 2HDM masses $(M_1,M_2, \mu)$ and
profiles of the inert-sector parameters (``dark profiles",
$M_S,M_A,M_{\eta^\pm}$).
\item
Next we check if for a given choice of dark profile $(M_S,M_A,M_{\eta^\pm})$
and $(M_1,M_2, \mu)$, there exists a value of $m_\eta$
such that the model predicts the right order of magnitude for the dark matter
abundance.
\item 
Then we scan over the mixing
angles $(\alpha_1, \alpha_2,\alpha_3)$ of the 2HDM, requiring that:
\begin{itemize}
\item[$-$\ ]
The naturalness is alleviated for each Higgs boson
\begin{equation}
\frac{|\delta M^2|}{M^2} = |\alpha_t| \frac{\Lambda^2}{M^2} < D 
\end{equation}
where $M$ denotes a generic Higgs boson mass, $\delta M^2$ stands for the
top-quark contribution to the one-loop correction to $M^2$, $\alpha_t$ is a
calculable coefficient in terms of the mixing angles etc.  The cut-off should
be chosen to be in the TeV region, e.g.\ modestly $\Lambda\simeq 2\tev$. The
fine-tuning parameter $D$ is to be chosen according to our aesthetic
standards.
\item[$-$\ ] Based on the experience gained from \cite{Barbieri:2006dq} we
restrict the scan to heavy ($\geq 100\gev$) Higgs bosons in the 2HDM sector
(so that the little hierarchy problem could be that way reduced, that would be
an analog of the heavy SM Higgs of \cite{Barbieri:2006dq}). It is worth noting
that here some possible tension between parameters emerges. It may appear as
we increase Higgs masses in the 2HDM sector trying to retain small quartic
constants. Even though the masses could be raised by increasing $\mu^2\sim
m_{12}^2$ in the potential, nevertheless the mass of one scalar would still
remain $\sim v$ (see e.g.~\cite{Gunion:2002zf}). In order to increase its mass
some combination of quartic couplings in $V_{12}$ will have to be large,
therefore checking the unitarity in the 2HDM sector is essential to guarantee
that $\lambda_i$'s remain in a perturbative regime.
\item[$-$\ ] 
Remaining experimental constraints are satisfied.
\item[$-$\ ] CP is violated, i.e.\ $\alpha_i$ are far enough from their CPC
limits. We use the invariants $J_i$, $i=1,2,3$ \cite{Lavoura:1994fv} as a
measure of CP violation. Eventually we plot an average and/or maximum (with
respect to the mixing angles $\alpha_i$) for $|\Im J_1|$ and the electron 
electric dipole moment (EDM) to illustrate the strength of CP violation.
\end{itemize} 
\end{itemize}

%%%%%%%%%%%%%%%%%%%%%%%%%%%%%%%%%%%%%%%%%%%%%%%%%%%%%%%%%%%%%%%%%%%%%%%%%%%%%
\section{Theoretical constraints}
\label{Sec:theory-constraints}
\setcounter{equation}{0}
%%%%%%%%%%%%%%%%%%%%%%%%%%%%%%%%%%%%%%%%%%%%%%%%%%%%%%%%%%%%%%%%%%%%%%%%%%%%%

%%%%%%%%%%%%%%%%%%%%%%%%%%%%%%%%%%%%%%%%%%%%%%%%%%%%%%%%%%%%%%%%%%%%%%%%%%%%%
\subsection{The little hierarchy}
\label{Sec:little-hierarchy}
%%%%%%%%%%%%%%%%%%%%%%%%%%%%%%%%%%%%%%%%%%%%%%%%%%%%%%%%%%%%%%%%%%%%%%%%%%%%%

As an order-of-magnitude estimate for radiative corrections to neutral Higgs
boson masses, we consider the contributions that arise from top-quark loops:
\begin{equation}
\delta M_{j}^2 = -\frac{3m_t^2}{4\pi^2v^2}\Lambda_{j}^2 
(a_j^2+\tilde a_j^2) \quad\text{for}\quad j=1,2,3
\label{hcor}
\end{equation}
where $a_j$ and $\tilde a_j$ are defined in (3.21) of \cite{El Kaffas:2006nt}:
\begin{equation}
a_j\equiv \frac{R_{j2}}{\sb}, \quad  \tilde a_j\equiv -\frac{\cb R_{j3}}{\sb}.
\label{aadef}
\end{equation}
Similarly, for the charged Higgs particles we find
\begin{equation}
\delta M_{H^\pm}^2=-\frac{3m_t^2}{4\pi^2v^2}\Lambda_{H^\pm}^2\cot^2\beta.
\end{equation}
Since the inert doublet does not couple to fermions there is no
hierarchy problem for $S$ and $A$. 

We adopt the following simple condition
\begin{equation} \label{Eq:little-hierarchy}
\frac{|\delta M_{j}^2|}{M_{j}^2} < D, \quad
\frac{|\delta M_{H^\pm}^2|}{M_{H^\pm}^2} < D
\end{equation} 
with the amount of fine tuning parametrized by $D$. For the resulting cut-off
we will chose 
\begin{equation} \label{Eq:cap_lambda}
\Lambda=\min(\Lambda_{j},\Lambda_{H^\pm}).
\end{equation}
This quantity $\Lambda$ will in general be most constrained by the value of
$M_1$ (unless when $a_1^2+\tilde a_1^2$ is small). For an alleviation of the
hierarchy problem, we would like to have $\Lambda$ ``large'' compared to the
Higgs masses.

%%%%%%%%%%%%%%%%%%%%%%%%%%%%%%%%%%%%%%%%%%%%%%%%%%%%%%%%%%%%%%%%%%%%%%%%%%%%%
\subsection{Perturbativity and Unitarity}
%%%%%%%%%%%%%%%%%%%%%%%%%%%%%%%%%%%%%%%%%%%%%%%%%%%%%%%%%%%%%%%%%%%%%%%%%%%%%

To preserve perturbativity we shall impose the following conditions on quartic
and Yukawa couplings of neutral and charged Higgs bosons
\begin{equation}
\lambda_i, \frac{\sqrt{2} m_t}{v}|a_j|, \frac{\sqrt{2} m_t}{v}|\tilde a_j|, 
\frac{ m_t}{\sqrt{2}v}\cot\beta, \lambda_a, \lambda_b, \lambda_c < 4 \pi.
\label{pert}
\end{equation}
Since we will consider $\tan\beta \geq 0.5$, the last three
conditions will always be satisfied.
We also impose unitarity on the Higgs-Higgs scattering amplitudes
\cite{Kanemura:1993hm,Akeroyd:2000wc,Ginzburg:2003fe}.
%%%%%%%%%%%%%%%%%%%%%%%%%%%%%%%%%%%%%%%%%%%%%%%%%%%%%%%%%%%%%%%%%%%%%%%%%%%%%
\subsection{CP Violation}
\label{subsec:CP-Violation}
%%%%%%%%%%%%%%%%%%%%%%%%%%%%%%%%%%%%%%%%%%%%%%%%%%%%%%%%%%%%%%%%%%%%%%%%%%%%%
Since our intention here is to outline a model which would possess CP
violation in the Higgs potential we shall discuss this issue in more detail.
The magnitude of CP violation can be quantified in terms of the invariants
introduced by Lavoura and Silva \cite{Lavoura:1994fv}.  However here we prefer
to adopt the more general, basis-independent approach of Gunion and Haber and
calculate the invariants $J_1$, $J_2$ and $J_3$ of \cite{Gunion:2005ja}. They
state (Theorem 4) that the Higgs sector is CP-conserving if and only if all
$J_i$ are real. The calculations of these quantities are straightforward, we
end up with the following result, valid for our choice of basis:
\begin{eqnarray}
\Im J_1&=&-\frac{v_1^2v_2^2}{v^4}(\lambda_1-\lambda_2)\Im \lambda_5,\\
\Im J_2&=&-\frac{v_1^2v_2^2}{v^8}
\left[\left((\lambda_1-\lambda_3-\lambda_4)^2-|\lambda_5|^2\right) v_1^4
+2(\lambda_1-\lambda_2) \Re \lambda_5 v_1^2v_2^2\right.\nonumber\\
&&\hspace*{1.5cm}\left.
-\left((\lambda_2-\lambda_3-\lambda_4)^2-|\lambda_5|^2\right) v_2^4\right]
\Im \lambda_5,\\
\Im J_3&=&\frac{v_1^2v_2^2}{v^4}(\lambda_1-\lambda_2)
(\lambda_1+\lambda_2+2\lambda_4)\Im \lambda_5.
\end{eqnarray}
We note that since we have chosen a basis with real v.e.v.'s, there is no CP
violation when $\Im\lambda_5=0$ (it should be realized that $\Im m_{12}^2$ and
$\Im \lambda_5$ are not independent here, see (\ref{m12})). Then an
interesting question arises: Is it possible, for $\Im\lambda_5\neq0$ to have
no CP violation? It turns out that the answer is ``yes'', as will be discussed
in the following.

The simultaneous vanishing of the three $\Im J_i$ implies that CP is
conserved. This can happen for five distinct cases:

\begin{itemize}
\item {\bf Case A:} $\Im\lambda_5\neq0$ with $v_1=0$ ($\tan\beta\to\infty$).
\item {\bf Case B:} $\Im\lambda_5\neq0$ with $v_2=0$ ($\tan\beta=0$).
\item {\bf Case C:} $\Im \lambda_5=0$. This corresponds to ${\cal M}^2$ (see
(\ref{Eq:cal-M})) being block diagonal, and the diagonalization performed in
terms of only {\it one} rotation angle.
\item {\bf Case D:} $\Im\lambda_5\neq0$ with $\lambda_1=\lambda_2$ and
$v_1=v_2$.
\item {\bf Case E:} $\Im\lambda_5\neq0$ with $\lambda_1=\lambda_2$,
$v_1\neq v_2$ and $(\lambda_1-\lambda_3-\lambda_4)^2 =|\lambda_5|^2$.
\end{itemize}

Some comments are here in order (details are discussed in Appendix~B):
\begin{itemize}
\item
When all three masses are degenerate, CP is conserved because
$\Im\lambda_5=0$.  This corresponds to Case C.

\item
When there is only partial mass degeneracy, $M_1=M_2<M_3$ or $M_1<M_2=M_3$,
there are instances of CP conservation corresponding to cases C, D and E.

\item
There are also instances of CP conservation in the mass non-degenerate case
$M_1<M_2<M_3$ corresponding to cases C, D and E.
\end{itemize}
The above discussion shows that, in terms of our input parameters ($\tan
\beta$, $M_1$, $M_2$, $M_{H^\pm}$ and $\mu$, together with $(\alpha_1,
\alpha_2,\alpha_3)$) there exist various non-trivial locations such that CP is
conserved even though $\Im\lambda_5\neq 0$ (the case of $\Im\lambda_5 = 0$ is
relatively obvious). Since our intention is to build a model that allows for a
substantial amount of CP violation, we would like to show regions of parameter space
where that indeed happens.  However, in light of the above discussion, the
determination of such locations can not easily be performed analytically.

The chance for successful electroweak baryogenesis is the crucial motivation 
for our discussion of CP violation. However, without a dedicated analysis of
baryogenesis (which is beyond the scope of this project) 
it is hard to estimate the amount of CP violation that is necessary. Therefore, we have
adopted the following strategy to illustrate the strength of CP violation {\it which is available}
in the model.
We plot both the electron electric dipole moment $d_e$ and the invariant $|\Im J_1|$
(which are physical quantities) in the region of the parameter space allowed by all the other
constraints. In order to estimate the amount of potential 
CP violation we show both averaged and maximal values of $d_e$ and $|\Im J_1|$ 
(the choice of $J_1$ (as opposed to $J_2$ and $J_3$) is arbitrary, 
however we recall that it is 
sufficient for CP violation to have just one of the $J_i$ complex). Large splitting between 
averaged and maximal values indicates the potential for CP violation hidden
in the appropriate choice of mixing angles $\alpha_i$. Of course, in a 
realistic situation (having the prediction for electroweak baryogenesis within the model) 
we would need to have $d_e$ below the experimental upper limit and nevertheless
enough CP violation for successful baryogenesis. 

In Appendix~B, in Figs.~\ref{alphas-075-110-112-100-300-200} and
\ref{alphas-077-110-112-400-500-400}, we show how allowed regions in
the $(\alpha_1, \alpha_2,\alpha_3)$ space are distributed.

%%%%%%%%%%%%%%%%%%%%%%%%%%%%%%%%%%%%%%%%%%%%%%%%%%%%%%%%%%%%%%%%%%%%%%%%%%%%%
\section{Experimental constraints}
\setcounter{equation}{0}
\label{Sec:exp-constraints}
%%%%%%%%%%%%%%%%%%%%%%%%%%%%%%%%%%%%%%%%%%%%%%%%%%%%%%%%%%%%%%%%%%%%%%%%%%%%%
We here review various experimental constraints that will be imposed on
the model.

%%%%%%%%%%%%%%%%%%%%%%%%%%%%%%%%
{\boldmath $T$ and $S$:}
We adopt the results from \cite{Grimus:2007if,Grimus:2008nb}
in order to calculate $T$ and $S$ within our model.  
Since for the Higgs fields we will use a basis in which  
only $\Phi_1$ has non-zero vev (so called Higgs basis)
some necessary transformations must be performed, 
see Appendix~C for details. For the
model discussed here the rotation matrix $O$ defined by Eq.~(59) of
\cite{Grimus:2007if} reads
\begin{equation} \label{Eq:Grimus-O}
O=\left(
\begin{array}{ccccc}
O_{11}&O_{12}&O_{13}&0&0\\
O_{21}&O_{22}&O_{23}&0&0\\
O_{31}&O_{32}&O_{33}&0&0\\
0&0&0&1&0\\
0&0&0&0&1
\end{array}
\right)
\end{equation}
and the mixing matrix $V$ of that paper becomes
\begin{equation}
V=\left(
\begin{array}{cccccc}
i&O_{11}&O_{12}&O_{13}&0&0\\
0&O_{21}+iO_{31}&O_{22}+iO_{32}&O_{23}+iO_{33}&0&0\\
0&0&0&0&1&i
\end{array}
\right),
\end{equation}
with $U=\boldsymbol{1}$. 
Therefore, $U^\dagger V=V$, and
\begin{equation}
V^\dagger V = \left(
\begin{array}{cccccc}
1&-iO_{11}&-iO_{12}&-iO_{13}&0&0\\
iO_{11}&1&iO_{13}&-iO_{12}&0&0\\
iO_{12}&-iO_{13}&1&iO_{11}&0&0\\
iO_{13}&iO_{12}&-iO_{11}&1&0&0\\
0&0&0&0&1&i\\
0&0&0&0&-i&1
\end{array}
\right)
\end{equation}
Since this is block diagonal, the contribution to $T$ (and to $S$)
from the inert doublet is additive (the inert fields must always appear
in pairs, there is no interference between the ``visible'' and the inert
sector at the one-loop order):
\begin{equation}
T=T_\text{2HDM} + \frac{1}{16\pi\sin^2\theta_W m_W^2}
\left[F(M_{\eta^\pm}^2,M_S^2) + F(M_{\eta^\pm}^2,M_A^2) - F(M_A^2,M_S^2)\right]
\label{drho}
\end{equation}
where $\alpha_\text{e.m.}T_\text{2HDM}=\Delta\rho_\text{2HDM}$ is given by
(63) of \cite{Grimus:2007if}. In our case, the matrices (\ref{Umat}) and
(\ref{Vmat}) should be adopted. Similarly, $S$ can be obtained from the
results given in \cite{Grimus:2008nb}. We impose the bounds $|\Delta T|<0.10$,
$|\Delta S|<0.10$ \cite{Amsler:2008zz}, at the 1-$\sigma$ level.

We note that $T$, which is our main concern, gets a positive contribution from
a splitting between the masses of charged and neutral Higgs bosons, whereas a
pair of neutral ones gives a negative contribution.  In fact, since the
function $F$ is symmetric in its two arguments, these two opposite-sign
contributions cancel in the limit when the charged boson is degenerate with
either of the two neutral ones.

%%%%%%%%%%%%%%%%%%%%%%%%%%%%%%%%
{\boldmath $B_0-\bar{B}_0$} mixing:
Due to the possibility of charged-Higgs exchange, in addition to $W^\pm$
exchange, the $B_0-\bar{B}_0$ mixing constraint excludes low values of
$\tan\beta$ and low values of $M_{H^\pm}$
\cite{Abbott:1979dt,Inami:1980fz,Urban:1997gw}.  Here we follow the procedure
of \cite{WahabElKaffas:2007xd}.

%%%%%%%%%%%%%%%%%%%%%%%%%%%%%%%%
{\boldmath $B\to X_s \gamma$}:
The $b\to s \gamma$ transition may also proceed via charged-Higgs exchange, so
some regions of the parameter space with low values of $\tan\beta$ and
$M_{H^\pm}$ are excluded. The exact region of exclusion is sensitive to
higher-order QCD effects
\cite{Chetyrkin:1996vx,Borzumati:1998tg,Misiak:2004ew,Bauer:1997fe}, and
roughly excludes $M_{H^\pm}<300~\text{GeV}$.  Again, we follow the approach of
\cite{WahabElKaffas:2007xd}.

%%%%%%%%%%%%%%%%%%%%%%%%%%%%%%%%
{\boldmath $B\to \tau \bar\nu_\tau X$}:
The charged Higgs contribution may substantially modify the branching ratio
for $B\to \tau \bar\nu_\tau X$ \cite{Krawczyk:1987zj}.  The measurement
\cite{Abbiendi:2001fi} of ${\cal B}(B\to \tau \bar\nu_\tau X)$ leads to the
following constraint
\begin{equation}
\frac{\tan\beta}{M_{H^\pm}} <  0.53 \gev^{-1}
\end{equation}
at $95\%$ CL. This is in fact a very weak constraint.
A more recent measurement gives ${\cal B}(B^-\to\tau\bar\nu_\tau)
=(1.79\pm0.71)\times10^{-4}$ \cite{Ikado:2006un}, where we have added in
quadrature symmetrized statistical and systematic errors. With the SM
prediction of $(1.59\pm0.40)\times10^{-4}$,
\begin{equation}
r_{H\,\text{exp}}=\frac{{\cal B}(B^-\to\tau\bar\nu_\tau)}
{{\cal B}(B^-\to\tau\bar\nu_\tau)_\text{SM}}
=1.13\pm0.53.
\end{equation}
Within the framework of the 2HDM, one finds \cite{Hou:1992sy}
\begin{equation}
r_{H\,\text{2HDM}}=\biggl[1-\frac{m_B^2}{M_{H^\pm}^2}\,\tan^2\beta\biggr]^2.
\end{equation}
Then the data imply that two sectors at large values of $\tan\beta$ and low
values of $M_{H^\pm}$ are excluded.

%%%%%%%%%%%%%%%%%%%%%%%%%%%%%%%%
{\boldmath $B\to D\tau \bar\nu_\tau$}:
Measurements~\cite{Aubert:2007dsa} of the ratio 
\begin{equation}
R^\text{exp}
=\frac{{\cal B}(B\to D\tau\nu_\tau)}{{\cal B}(B\to D\ell\nu_\ell)},
\quad \ell=e,\mu,
\end{equation}
can also be used to constrain the coupling of the charged Higgs to the $\tau$,
more precisely $\tan\beta/M_{H^\pm}$.  It thus restricts large values of
$\tan\beta$ and low values of $M_{H^\pm}$ \cite{Nierste:2008qe}, in a region
of parameter space similar to the one following from $B\to \tau \bar\nu_\tau
X$, but is considerably stronger.

%%%%%%%%%%%%%%%%%%%%%%%%%%%%%%%%
{\bf LEP2 non-discovery:}
The non-discovery of a Higgs boson at LEP2 imposes a bound on how strongly the
lightest one can couple to the $Z$ and to $b\bar b$.  Useful results for this
constraint are available in table 27 of \cite{Abdallah:2004wy}. Adopting the
standard notation
\begin{equation}
\sigma_{Z(H\to b\bar b)}
=\sigma_{Z(H\to b\bar b)}^{SM}\times C^2_{Z(H\to b\bar b)}
\end{equation}
one can approximately parametrize the upper limit on $C^2_{Z(H\to b\bar b)}$
from the table as follows:
\begin{equation}
C^2_{Z(H\to b\bar b)} \leq \left\{
\begin{array}{lll} 
0.05 & \text{for} & 12\gev < M_A \leq 80 \gev,\\
0.1 & \text{for} & 80\gev < M_A \leq 90 \gev,\\
0.2 & \text{for} & 90\gev < M_A < 110 \gev.\\
\end{array}
\right.
\label{lep}
\end{equation}
For $e^+e^- \to ZH_1$ the coefficient $C^2_{Z(H\to b\bar b)}$ is given by
(4.3) of \cite{El Kaffas:2006nt}:
\begin{equation}
C^2_{Z(H_1\to b\bar b)} 
= \left(\cb R_{11}+\sb R_{12}\right)^2\frac{1}{\cb^2}\left(R_{11}^2
+\sb^2 R_{13}^2\right).
\label{c2def}
\end{equation}
Then the constraint (\ref{lep}) limits the parameter space. 

%%%%%%%%%%%%%%%%%%%%%%%%%%%%%%%%
{\boldmath $R_b$}:
The branching ratio for $Z\to b\bar b$ is also affected by Higgs exchange.
As noticed in \cite{El Kaffas:2006nt}, the contributions from neutral Higgs
bosons to $R_b$ are negligible, however, charged Higgs boson contributions, as
given by \cite{Denner:1991ie}, Eq.~(4.2), exclude low values of $\tan\beta$
and low $M_{H^\pm}$.  Experimentally $R_b \equiv \Gamma_{Z\to b\bar b} /
\Gamma_{Z\to {\rm had}} = 0.21629 \pm 0.00066$ \cite{Amsler:2008zz}.  It is
easy to see that the correction $\delta \Gamma_{Z\to b\bar b}$ implies the
following change for $R_b$:
\begin{equation}
\delta R_b=\frac{\delta \Gamma_{Z\to b\bar b}}{\Gamma_{Z\to {\rm had}}}(1-R_b)
\end{equation}
where $\Gamma_{Z\to {\rm had}}= (1.7444 \pm 0.0020)~\text{GeV}$
\cite{Amsler:2008zz}. Since $\delta \Gamma_{Z\to b\bar b}$ is known within the
2HDM~\cite{Denner:1991ie}, so is $\delta R_b$. 
We require
\begin{equation}
\delta R_b < 0.00066,
\end{equation}
corresponding to $\delta\Gamma_{Z\to b\bar b}=1.47~\text{MeV}$ at the 1-$\sigma$
level.

%%%%%%%%%%%%%%%%%%%%%%%%%%%%%%%%
{\bf Muon anomalous magnetic moment:}
Since here we are considering heavy Higgs bosons ($M_i \gsim 100 \gev$)
therefore, according to \cite{Cheung:2003pw,WahabElKaffas:2007xd}, the 2HDM
contribution to the muon anomalous magnetic moment is negligible even for
$\tan\beta$ as large as $\sim 40$.

%%%%%%%%%%%%%%%%%%%%%%%%%%%%%%%%
{\bf Electron electric dipole moment:}
The bounds on electric dipole moments constrain the allowed amount of CP
violation of the model. We adopt the bound \cite{Regan:2002ta} (see also
\cite{Pilaftsis:2002fe}):
\begin{equation}
|d_e|\lsim1\times10^{-27} [e\,\text{cm}],
\end{equation}
at the 1-$\sigma$ level.
The contribution due to neutral Higgs exchange\footnote{Neglecting the CKM
mixing, the charged Higgs contribution to $d_e$ vanishes in the 2HDM
 (with softly broken $Z_2$) up to two loops \cite{BowserChao:1997bb}.}, via the
two-loop Barr--Zee effect \cite{Barr:1990vd}, is given by Eq.~(3.2) of
\cite{Pilaftsis:2002fe} in terms of the neutral-sector mixing matrix $O$,
defined in \cite{Pilaftsis:1999qt}, and related to our $R$ via
\begin{equation}
\begin{pmatrix}
O_{11} & O_{12} & O_{13} \\
O_{21} & O_{22} & O_{23} \\
O_{31} & O_{32} & O_{33}
\end{pmatrix}
=
\begin{pmatrix}
R_{33} & R_{23} & R_{13} \\
R_{31} & R_{21} & R_{11} \\
R_{32} & R_{22} & R_{12}
\end{pmatrix}.
\end{equation}
This rotation matrix $O$ should not be confused with the one appearing in 
Eq.~(\ref{Eq:Grimus-O}). 
%%%%%%%%%%%%%%%%%%%%%%%%%%%%%%%%%%%%%%%%%%%%%%%%%%%%%%%%%%%%%%%%%%%%%%%%%%%%%
\section{Results}
\label{Sec:results}
\setcounter{equation}{0}
%%%%%%%%%%%%%%%%%%%%%%%%%%%%%%%%%%%%%%%%%%%%%%%%%%%%%%%%%%%%%%%%%%%%%%%%%%%%%
Subject to the limitations discussed above, we may now scan over the parameter
space, imposing the constraint
\begin{equation} \label{Eq:exp-constraints}
\chi^2=\sum\chi_i^2<5.99,\quad 95\%\text{C.L.},
\end{equation}
as appropriate for identifying allowed regions in two dimensions,
($\tan\beta,M_{H^\pm}$).  The sum runs over all the experimental constraints
discussed above. 

Our results will be given in terms of contour plots of various quantities of
interest.  Regions corresponding to values being confined within certain
intervals are indicated by a colour coding as indicated. The external
contour shows the maximal region consistent with the experimental and
theoretical constraints we adopt (i.e., without the naturality condition
imposed on $\Lambda/\sqrt{D}$, unless explicitly stated).

The model discussed here contains many parameters (masses, mixing angles, etc.).
In projecting down our results to a lower-dimensional space, 
we have decided to favour the more ``physical" parameters $\tan\beta$ and $M_{H^\pm}$ 
together with neutral scalar masses (and $\mu$). 
In this section, most of the plots show (for fixed inert masses, $M_S, M_A, M_{\eta^\pm}$, 
fixed 2HDM neutral masses, $M_1,M_2$ and $\mu$) allowed regions in the $(\tan\beta,M_{H^\pm})$
space. The remaining parameters, the neutral-Higgs-sector mixing angles, $\alpha_{1}$,
$\alpha_{2}$, and $\alpha_{3}$, which are not specified in those plots, have been averaged over,
or a maximum has been extracted.
Thus, for each allowed point in ($\tan\beta,M_{H^\pm}$) there exist $\alpha$'s such that all constraints
are satisfied.

%%%%%%%%%%%%%%%%%%%%%%%%%%%%%%%%%%%%%%%%%%%%%%%%%%%%%%%%%%%%%%%%%%%%%%%%%%
\begin{figure}[htb]
%\vspace*{-2.0cm}
\centerline{
\includegraphics[width=15.5cm,angle=0]{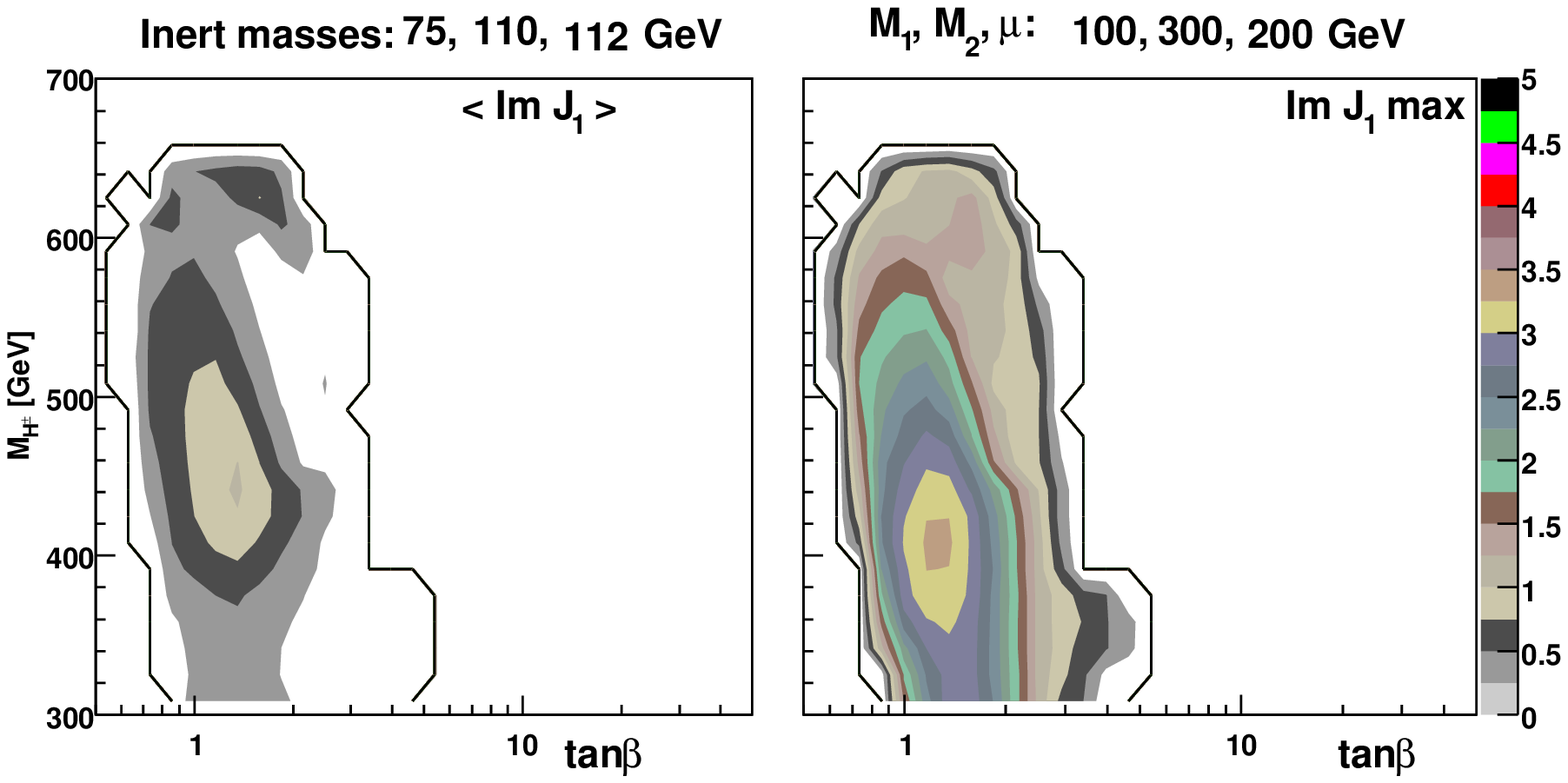}}
\centerline{
\includegraphics[width=15.5cm,angle=0]{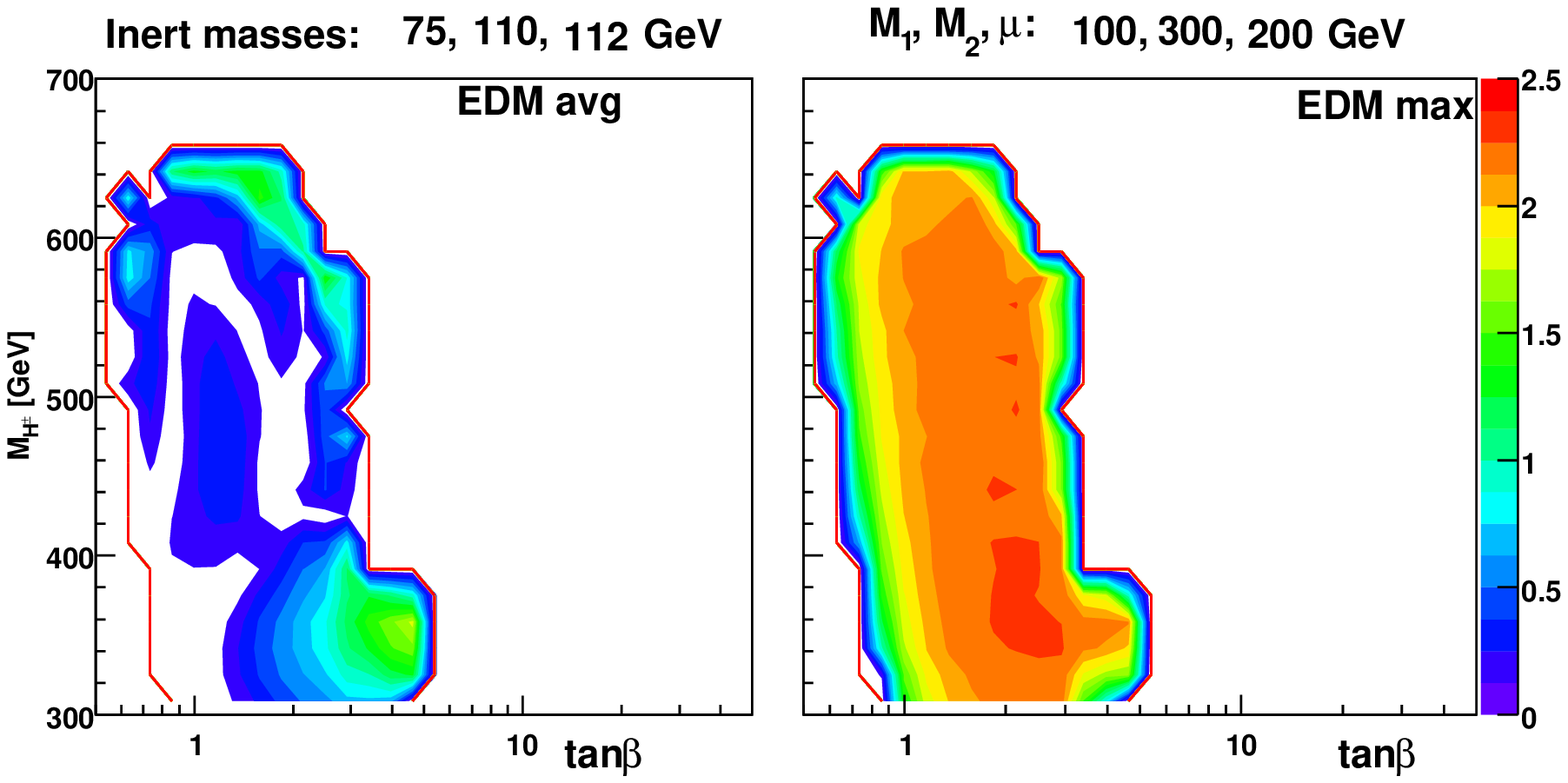}
}
\caption{\label{075-110-112-100-300-200} Top panels: Relative amount of CP
violation. Bottom panels: Electron electric dipole moment
in units [$e$ $10^{-27}~\text{cm}$].
Left: average; Right: maximum value. Inert-sector
masses: $(75, 110, 112)$~GeV; 2HDM-sector masses: ($M_1$, $M_2$, $\mu$) =
(100, 300, 200)~GeV.}
\end{figure}
%%%%%%%%%%%%%%%%%%%%%%%%%%%%%%%%%%%%%%%%%%%%%%%%%%%%%%%%%%%%%%%%%%%%%%%%%%

%%%%%%%%%%%%%%%%%%%%%%%%%%%%%%%%%%%%%%%%%%%%%%%%%%%%%%%%%%%%%%%%%%%%%%%%%%%%%
\subsection{Light Dark Matter particle}
\label{subsec:light-DM}
%%%%%%%%%%%%%%%%%%%%%%%%%%%%%%%%%%%%%%%%%%%%%%%%%%%%%%%%%%%%%%%%%%%%%%%%%%%%%

We first consider the set of ``dark profiles'' for which the DM particle is
light, at $\sim75$~GeV, see Eq.~(\ref{Eq:light-profiles}).  These profiles may
be compared with the parameters considered by Lopez Honorez et al.\
\cite{LopezHonorez:2006gr}, in their Fig.~5, where it was shown that the
correct amount of dark matter may be obtained with dark-matter mass of the
order of 50--80~GeV, for splittings of the order 10--50~GeV.

%%%%%%%%%%%%%%%%%%%%%%%%%%%%%%%%%%%%%%%%%%%%%
\begin{table}[htb]
\begin{center}
\begin{tabular}{|c|c|c|c|c|c|}
\hline
 & DP1 & DP1' & DP3 & DP4, DP5 \\
$\Omega h^2$
 & (75,110,112) & (77,110,112) & (75,110,85) & (100,110,115)
 \\
\hline
A (100,300,200)& OK & $<0.07$ & $<0.09$ & $<0.01$ \\
B (200,400,400) & $>0.24$ & OK & OK & $<0.01$ 
\\
C (400,500,400) & $>0.24$ & $0.08-0.09$ & $\lsim0.10$ & $<0.01$ 
\\
\hline
\end{tabular}
\end{center}
\caption{\label{Table:allowed0} Dark-matter density $\Omega h^2$ for different
``dark profiles'', DP1, DP1', DP3, DP4, DP5 with mass parameters ($M_S, M_A,
M_{\eta^\pm}$), and different mass-parameter sets A--C with mass parameters
($M_1, M_2, \mu$), all in GeV, see Sec.~\ref{Sec:benchmarks}.}
\end{table}
%%%%%%%%%%%%%%%%%%%%%%%%%%%%%%%%%%%%%%%%%%%%%

In Table~\ref{Table:allowed0} we summarize the results for the DM density
$\Omega h^2$ for the different combinations of dark profiles of
Eq.~(\ref{Eq:light-profiles}), and non-inert-sector parameters of
Eq.~(\ref{Eq:non-inert-masses}).  These are obtained from running {\tt
micrOMEGAs} \cite{Belanger:2006is,Belanger:2008sj}, as described in
Sec.~\ref{Sec:benchmarks}.  For the quartic couplings in the inert sector
(denoted here by $\lambda_\eta/2$), we use the value 0.1 (the amount of dark
matter is practically independent of this coupling
\cite{LopezHonorez:2006gr}).  Here, ``OK'' means that a value within $3\sigma$
of the WMAP value $0.1131 \pm 0.0034$ \cite{Bennett:2003bz}, 
can be found for a suitable choice of $m_\eta$. For these profiles of light DM particles, 
typically values $m_\eta\sim30-50~\text{GeV}$ are required.

Profile~1' corresponds to the allowed horizontal band in Fig.~9 of Lundstr\"om
et al.\ \cite{Lundstrom:2008ai}, in the sense that we find solutions for
Sets~B and C, corresponding to heavier non-inert Higgs particles.  This band
is rather narrow: for $M_S=75~\text{GeV}$ we find too much dark matter, at
$M_S=79~\text{GeV}$ too little. It appears shifted by a couple of GeV, with
respect to the results of \cite{Lundstrom:2008ai}, presumably due to the use
of a different DM code, {\tt micrOMEGAs}
\cite{Belanger:2006is,Belanger:2008sj} vs. {\sc DarkSusy}
\cite{Gondolo:2004sc}.

All these profiles (except Profile~3) have the charged particle heavier than
both the neutral ones, in order that the inert sector makes a positive
contribution to $T$ (see Sec.~\ref{Sec:exp-constraints}). This, in turn,
allows for a higher value of the ``ordinary'' neutral Higgs particle
\cite{Barbieri:2006dq} that enters the electroweak fits.

Profiles~4 and 5 give $\Omega h^2\leq0.01$ (consistent with Figs.~8 and 9 of
\cite{Lundstrom:2008ai}) and will not be considered any further.

{\bf Dark profile~1.}
We start discussing the case $(M_S,M_A,M_{\eta^\pm})=(75,110,112)$~GeV.  Among
the three sets of mass parameters considered for the visible sector, only
Set~A ($M_1$, $M_2$, $\mu$) = (100, 300, 200)~GeV gives a reasonable value of
$\Omega h^2$.  With $m_\eta$ of the order 45--55~GeV we obtain $\Omega h^2
\simeq 0.10-0.13$. Higher values of $m_\eta$ yield too high values of $\Omega
h^2$.  For Profile~1 and Sets B and C, the value is too high, $\Omega
h^2\geq0.24$.\footnote{For Sets B and C the lightest 2HDM Higgs is relatively
heavy so that $SS$ annihilation via an intermediate $H_1$ is suppressed, and
too much dark matter would survive.}  The allowed region is shown in
Fig.~\ref{075-110-112-100-300-200}.  The figure is obtained by scanning over
$\tan\beta$ from 0.5 to 50, and over $M_{H^\pm}$ from 300~GeV (barely above
the $B\to X_s \gamma$ cut-off) to 700~GeV. For each point in the
$(\tan\beta,M_{H^\pm})$ plane, a scan over mixing angles
$(\alpha_1,\alpha_2,\alpha_3)$ is performed, analyzing all models compliant
with the constraints, showing contours of the CP-violating quantity $|\Im J_1|$
and the electron electric dipole moment (EDM).  These two quantities provide
two different measures of the amount of CP violation. The outer contours
delineate the regions within which consistent solutions are found for one or
more sets of mixing angles $\alpha_i$.
The averages presented in the left panels are obtained by averaging over 
all sets of $(\alpha_1,\alpha_2,\alpha_3)$ for which the experimental constraints
are satisfied according to Eq.~(\ref{Eq:exp-constraints}). Similarly,
the maxima in the right panels correspond to maxima of absolute values,
obtained from the same scans over allowed sets of $(\alpha_1,\alpha_2,\alpha_3)$.

In the upper part of the figure we display $|\Im J_1|$, while the lower one
shows the electron electric dipole moment, $|d_e|$. In the left panels
averages over allowed sets of $\alpha_i$ are displayed, while the right ones
show extrema. There is no obvious correlation between $|d_e|$ and $|\Im J_1|$,
other than both having maxima in the interior of the allowed region.

In Appendix~B we discuss the correlations of angles $\alpha_i$ for which
viable solutions are found. These are shown separately for ``small
$\tan\beta$'' and ``large $\tan\beta$''. 
In general, it is easier to accommodate CP violation at low values
of $\tan\beta$ than at higher values. For high values of $\tan\beta$,
the allowed values of these parameters tend to accumulate near the limits
where $H_2$ is odd under CP ($\alpha_2\simeq0$, $\alpha_3\simeq\pm\pi/2$).

For the parameters considered here, ($M_1$, $M_2$, $\mu$) = (100, 300,
200)~GeV, the viable models are constrained to $\tan\beta\sim 0.5 - 6$. The
cut-off at ``high'' $\tan\beta$ is mostly due to the unitarity constraint on
the neutral-Higgs sector. We also note a cut-off at
$M_{H^\pm}\sim650~\text{GeV}$. This, on the other hand, is due to the
electroweak constraints, $T$ in particular.

The fact that the {\it average} electric dipole moment is rather small
compared to the maximum value, means that large parts of the
$\tan\beta$-$M_{H^\pm}$ space would remain viable even if the experimental
constraint on $|d_e|$ should become significantly tightened.

%%%%%%%%%%%%%%%%%%%%%%%%%%%%%%%%%%%%%%%%%%%%%%%%%%%%%%%%%%%%%%%%%%%%%%%%%%
\begin{figure}[htb]
%\vspace*{-2.0cm}
\centerline{
\includegraphics[width=7.5cm,angle=0]{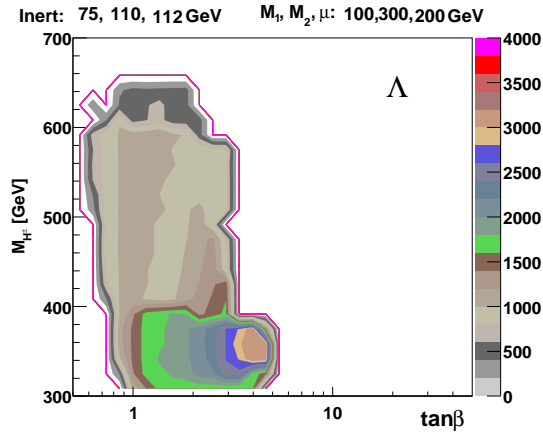}}
\caption{\label{cap_lam-075-110-112} Maximum reach in $\Lambda/\sqrt{D}$
[GeV]. Inert-sector masses: $(75, 110, 112)$~GeV; 2HDM-sector masses: ($M_1$,
$M_2$, $\mu$) = (100, 300, 200)~GeV.}
\end{figure}
%%%%%%%%%%%%%%%%%%%%%%%%%%%%%%%%%%%%%%%%%%%%%%%%%%%%%%%%%%%%%%%%%%%%%%%%%%

{\bf The little hierarchy.}
In this Fig.~\ref{075-110-112-100-300-200},
no constraint from the fine-tuning consideration in
Sec.~\ref{Sec:little-hierarchy} is imposed.
In Fig.~\ref{cap_lam-075-110-112} we plot the quantity $\Lambda/\sqrt{D}$ of
Eq.~(\ref{Eq:cap_lambda}), which by definition is minimized over the three
neutral Higgs particles and the charged one (for fixed mixing angles
$\alpha_i$). This is next maximized over the $\alpha_i$ (for fixed $\tan\beta$
and $M_{H^\pm}$) before being plotted in this figure.  For the cases
considered, we note that $\Lambda/\sqrt{D}$ reaches up to around 3~TeV.
The dominant reason why this varies with $\tan\beta$ and $M_{H^\pm}$ is
that the factor
\begin{equation}
\xi_j\equiv a_j^2+\tilde a_j^2
\end{equation}
of Eq.~(\ref{hcor}), for which solutions are allowed, varies. Specifically,
when the factor $\xi_1$ is small, $\Lambda/\sqrt{D}$ will be larger. This
happens when $\sin\alpha_1\cos\alpha_2$ and $\cos\beta\sin\alpha_2$ are both
small (cf.\ Appendix~B).

\begin{figure}[htb]
%\vspace*{-2.0cm}
\centerline{
\includegraphics[width=15.5cm,angle=0]{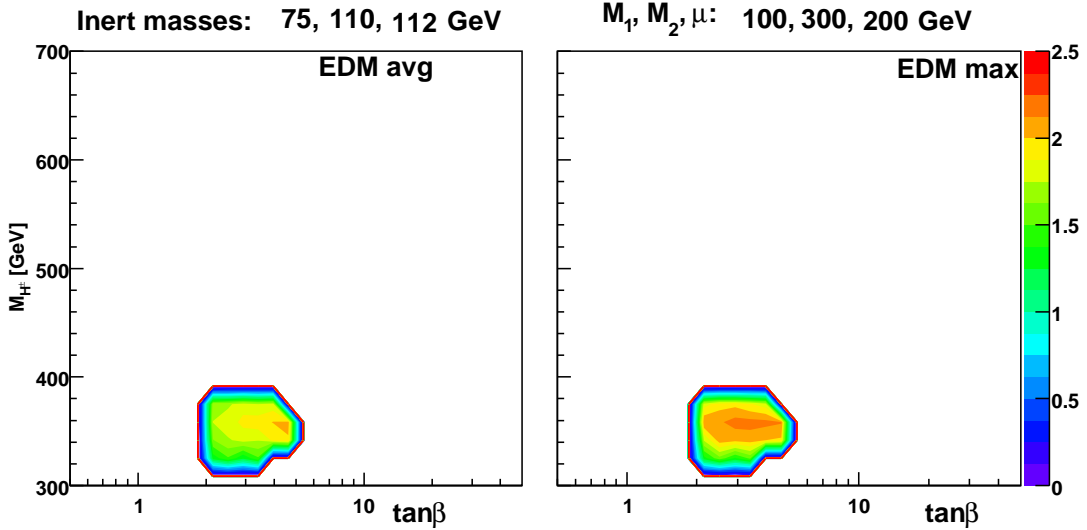}}
\caption{\label{075-110-112-lam=2} Allowed regions for
$\Lambda/\sqrt{D}=2~\text{TeV}$ (outer contours)
and electron electric dipole moment in units
[$e$ $10^{-27}~\text{cm}$].  Left: average; Right: maximum value. Inert-sector
masses: $(75, 110, 112)$~GeV; 2HDM-sector masses: ($M_1$, $M_2$, $\mu$) =
(100, 300, 200)~GeV.}
\end{figure}

Accordingly, imposing the constraint (\ref{Eq:little-hierarchy}), with
$\Lambda/\sqrt{D}=2~\text{TeV}$, we obtain the more restricted allowed regions
shown in Fig.~\ref{075-110-112-lam=2}. Compared with
Fig.~\ref{075-110-112-100-300-200} (where there is no constraint on
$\Lambda$), we see a dramatic reduction of the allowed parameter space. For
$M_1=100~\text{GeV}$, only a small region of $M_{H^\pm}\sim300-350~\text{GeV}$
and $\tan\beta\sim 2-6$ survives.  Naturally, this is dominantly caused by the
condition (\ref{Eq:little-hierarchy}) applied to $M_1$. 

%%%%%%%%%%%%%%%%%%%%%%%%%%%%%%%%%%%%%%%%%%%%%%%%%%%%%%%%%%%%%%%%%%%%%%%%%%
%%%%%%%%%%%%%%%%%%%%%%%%%%%%%%%%%%%%%%%%%%%%%%%%%%%%%%%%%%%%%%%%%%%%%%%%%%

\begin{figure}[htb]
%\vspace*{-2.0cm}
\centerline{
\includegraphics[width=15.5cm,angle=0]{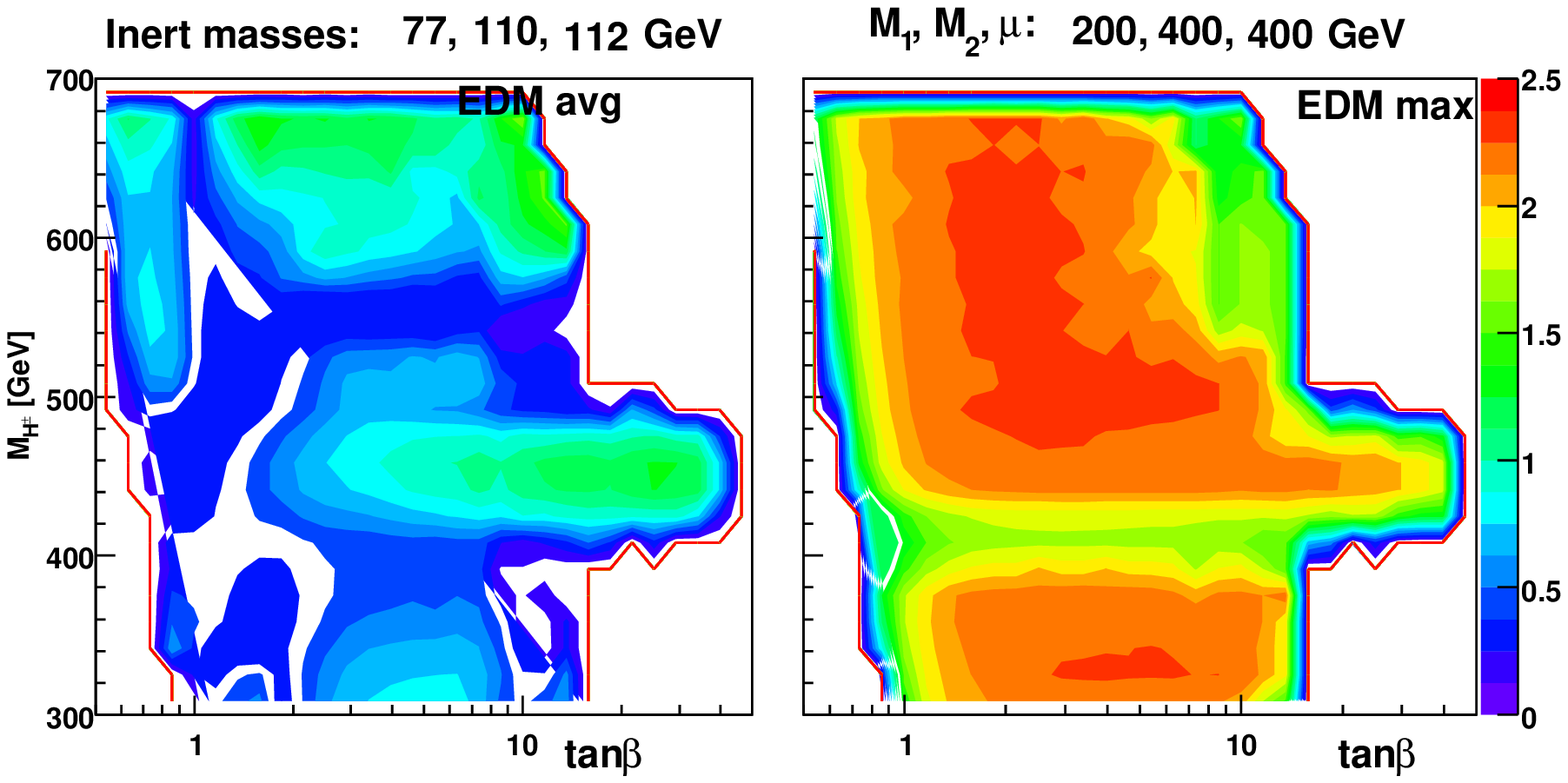}}
\centerline{
\includegraphics[width=15.5cm,angle=0]{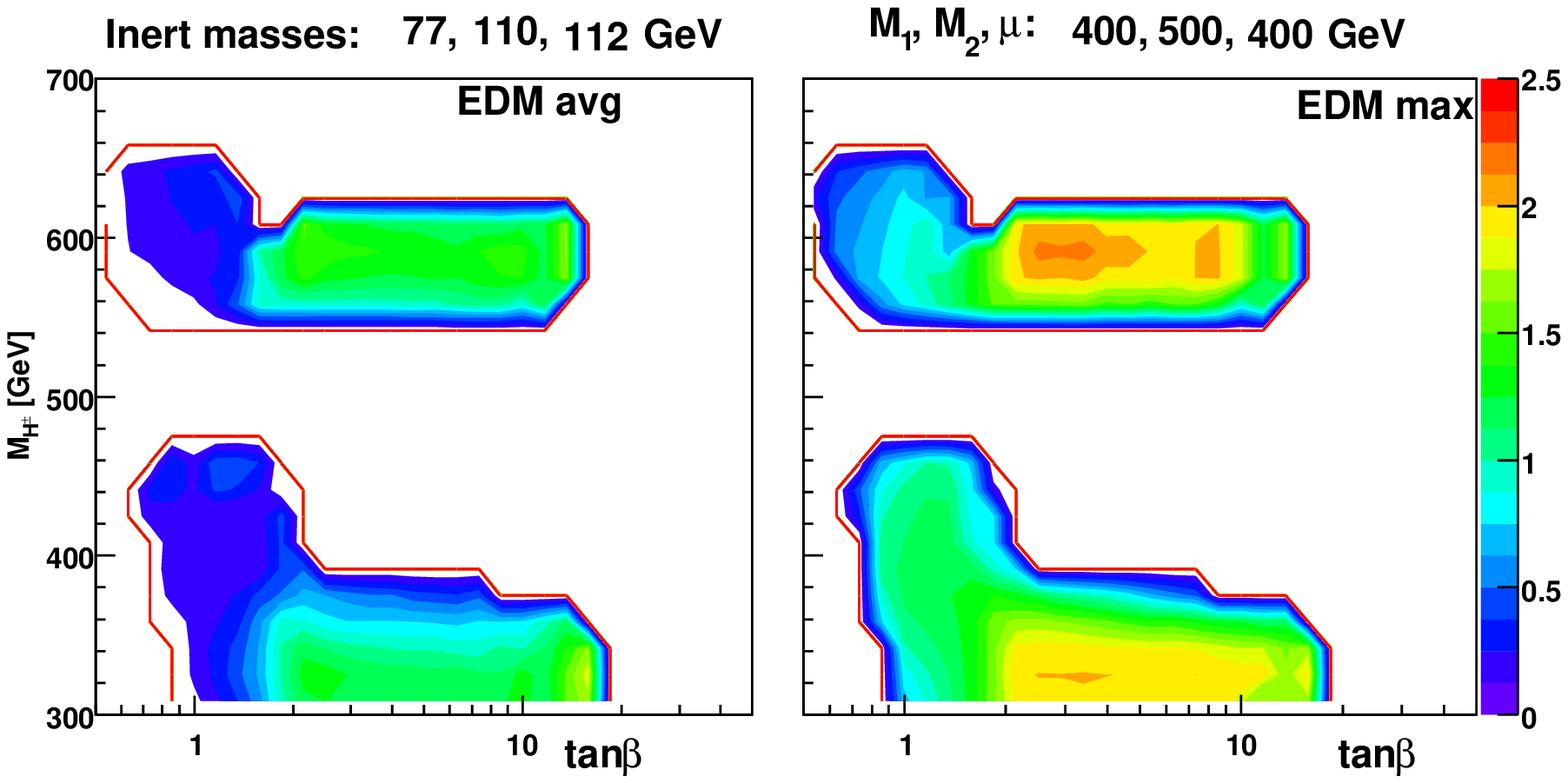}}
\caption{\label{edm-077-110-112} Electron electric dipole moment
in units [$e$ $10^{-27}~\text{cm}$].
Left: average; Right: maximum value.
 Inert-sector masses: $(77, 110, 112)$~GeV; 2HDM-sector masses: ($M_1$,
$M_2$, $\mu$) = (200, 400, 400)~GeV and (400, 500, 400)~GeV.}
\end{figure}

{\bf Dark profile~1'.}
This profile has a slightly higher value of $M_S$, and thus gives allowed
models also out to higher values of the non-inert Higgs mass, $M_1$,
corresponding to the ``horizontal band'' in Fig.~9 of
Ref.~\cite{Lundstrom:2008ai}.  The results for Sets~B and C are shown in
Fig.~\ref{edm-077-110-112}. As compared with profile~1, the heavier $S$ here
allows more annihilation to (off-shell) $WW$ pairs, and thus an acceptable
value of $\Omega h^2$ can be found.

For Set~B, there are solutions over most of the explored
$\tan\beta$--$M_{H^\pm}$ plane, for $M_{H^\pm}\sim400-500~\text{GeV}$ and
$M_2=\mu=400~\text{GeV}$ they reach all the way out to $\tan\beta\sim50$.  For
Set~C, on the other hand, the situation is different.  Apart from a region
around $\tan\beta\sim1-2$, solutions are only found within two widely
separated bands in the $\tan\beta$--$M_{H^\pm}$ plane. There is one band at
$M_{H^\pm}\lsim400~\text{GeV}$ and another at
$M_{H^\pm}\lsim550-650~\text{GeV}$, with a region of no solutions in-between.
This is clearly due to the interplay or partial cancellation of contributions
to $T$, positive from the charged-Higgs boson differing in mass from some
neutral ones, and negative from pairs of neutral ones having different masses.

%%%%%%%%%%%%%%%%%%%%%%%%%%%%%%%%%%%%%%%%%%%%%%%%%%%%%%%%%%%%%%%%%%%%%%%%%%
%%%%%%%%%%%%%%%%%%%%%%%%%%%%%%%%%%%%%%%%%%%%%%%%%%%%%%%%%%%%%%%%%%%%%%%%%%

{\bf Dark profile~2.}
In this case the allowed models are located in essentially the same part of
the $\tan\beta$--$M_{H^\pm}$ plane as for ``dark profile~1'', with similar
values for $|\Im J_1|$ and the electron electric dipole moment.

{\bf Dark profile~3.}
This profile differs from the previous ones in having the charged boson of the
inert sector lighter than the heavier neutral one.  With this modification
from Profile~1, we find acceptable solutions also for heavier Higgs bosons in
the non-inert sector (Sets~B and C).  The allowed regions in the
$\tan\beta$--$M_{H^\pm}$ plane are quite similar to those shown in
Fig.~\ref{edm-077-110-112} for Profile~1', and also the values of the electric
dipole moment are similar.

\begin{figure}[htb]
%\vspace*{-2.0cm}
\centerline{
\includegraphics[width=15.5cm,angle=0]{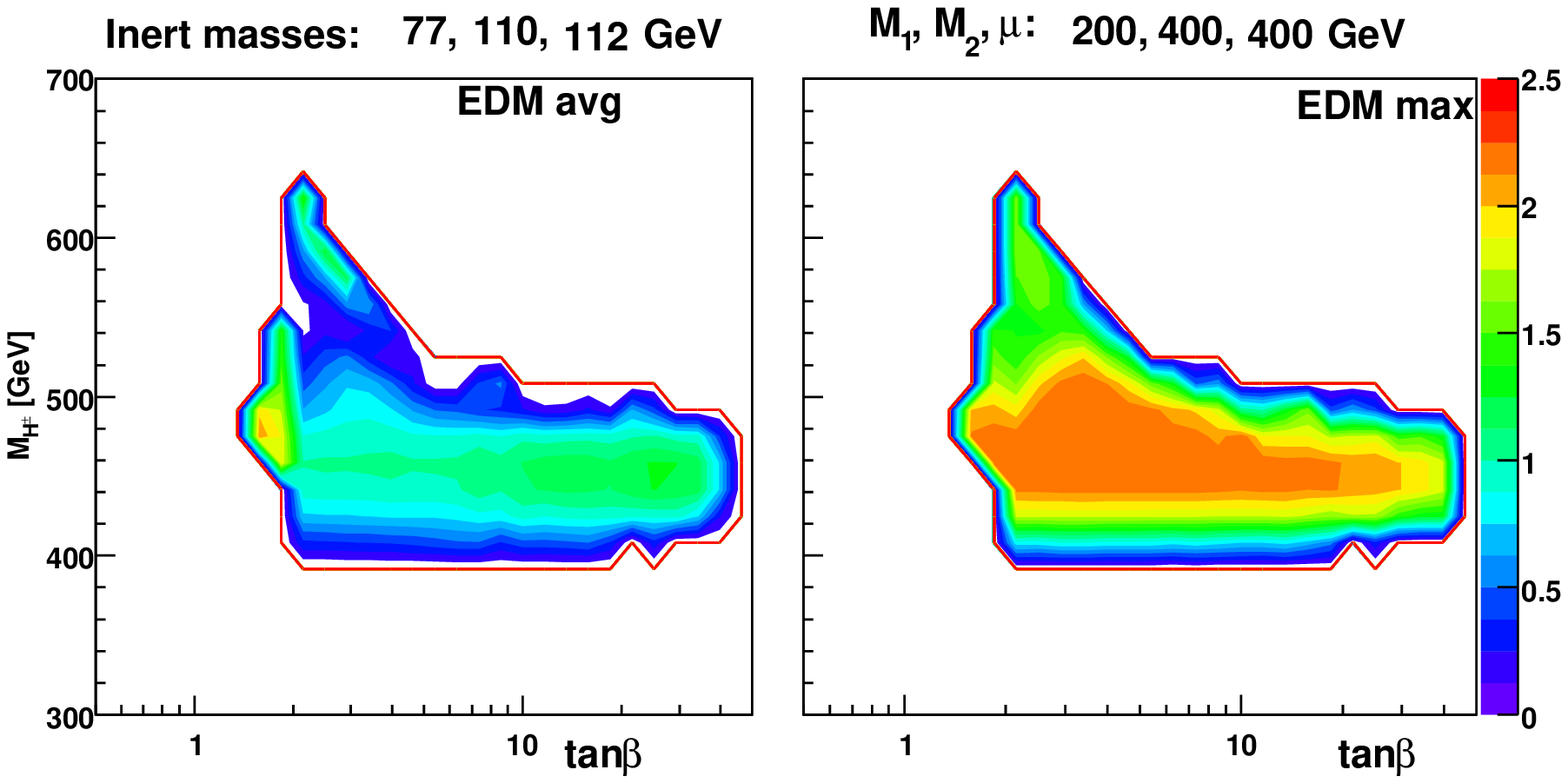}}
\centerline{
\includegraphics[width=15.5cm,angle=0]{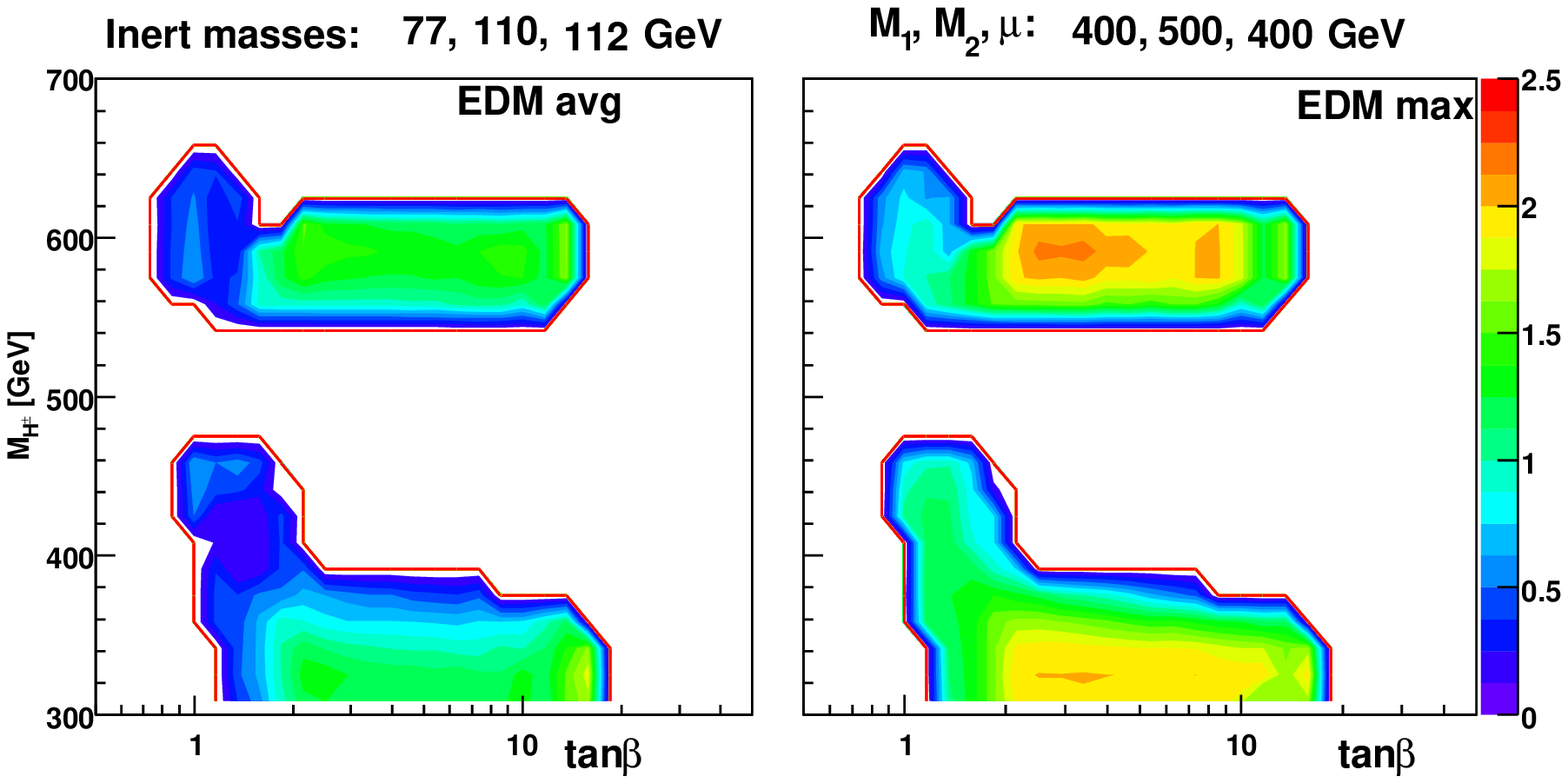}}
\caption{\label{077-110-112-lam=2} Allowed regions for
$\Lambda/\sqrt{D}=2~\text{TeV}$ (outer contours)
and electron electric dipole moment in units
[$e$ $10^{-27}~\text{cm}$].  Left: average; Right: maximum value. Inert-sector
masses: $(77, 110, 112)$~GeV; 2HDM-sector masses: ($M_1$, $M_2$, $\mu$) =
(200, 400, 400)~GeV and (400, 500, 400)~GeV.}
\end{figure}

{\bf The little hierarchy.}
If we here impose the constraint (\ref{Eq:little-hierarchy}), with
$\Lambda/\sqrt{D}=2~\text{TeV}$, there is a considerable reduction in the
allowed parameter space for Set~B, whereas for Set~C essentially the whole
parameter space survives, as shown in Fig.~\ref{077-110-112-lam=2} for
Profile~1'.
%%%%%%%%%%%%%%%%%%%%%%%%%%%%%%%%%%%%%%%%%%%%%%%%%%%%%%%%%%%%%%%%%%%%%%%%%%
%%%%%%%%%%%%%%%%%%%%%%%%%%%%%%%%%%%%%%%%%%%%%%%%%%%%%%%%%%%%%%%%%%%%%%%%%%

%%%%%%%%%%%%%%%%%%%%%%%%%%%%%%%%%%%%%%%%%%%%%%%%%%%%%%%%%%%%%%%%%%%%%%%%%%%%%
\subsection{Heavier Dark Matter particle}
\label{Subsec:Heavier_Dark Matter_particle}
%%%%%%%%%%%%%%%%%%%%%%%%%%%%%%%%%%%%%%%%%%%%%%%%%%%%%%%%%%%%%%%%%%%%%%%%%%%%%

The second set of ``dark profiles'' has a heavier DM particle. For
this case, it was shown (see Fig.~6 of \cite{LopezHonorez:2006gr}) that the
amount of dark matter can satisfy the WMAP constraint \cite{Bennett:2003bz}
provided the splitting in the inert sector is small.  We confirm that
finding and list in Table~\ref{Table:allowed-heavy} approximate values of
$\Omega h^2$ obtained from {\tt micrOMEGAs}
\cite{Belanger:2006is,Belanger:2008sj} with suitable choices of $m_\eta$. The
values vary quite a bit with the choice of this parameter.  As mentioned
above, the small mass splitting in this inert sector also leads to a small
(hence acceptable) modification of $T$.

%%%%%%%%%%%%%%%%%%%%%%%%%%%%%%%%%%%%%%%%%%%%%
\begin{table}[htb]
\begin{center}
\begin{tabular}{|c|c|c|c|c|c|}
\hline
 & DP11 & DP12 & DP13& DP14\\
 $\Omega h^2$
 & (500,501,502) & (600,601,602) & (800,802,804) & (1000,1002,1005)\\
\hline
A (100,300,200)& $\simeq 0.09$ 
& $\simeq 0.10$ 
& $\simeq 0.09$ 
& $\simeq 0.10$\\
B (200,400,400) & $\simeq 0.09$ 
& $\simeq 0.10$ 
& $\simeq 0.09$ 
& $\simeq 0.10$\\
C (400,500,400) & $\simeq 0.09$ 
& $\simeq 0.10$ 
& $\simeq 0.09$
& $\simeq 0.10$\\
\hline
\end{tabular}
\end{center}
\caption{\label{Table:allowed-heavy} DM density $\Omega h^2$ for
different ``dark profiles'', DP11, DP12, DP13, DP14 with mass parameters
($M_S, M_A, M_{\eta^\pm}$) and different 2HDM mass-parameter sets A--C, 
($M_1, M_2, \mu$), all in GeV, see
Sec.~\ref{Sec:benchmarks}.}
\end{table}
%%%%%%%%%%%%%%%%%%%%%%%%%%%%%%%%%%%%%%%%%%%%%

{\bf Dark profiles~11, 12, 13 and 14.}
For all these cases, we find solutions for all the three considered sets of
non-inert mass parameters of Eq.~(\ref{Eq:non-inert-masses}). The allowed
regions in the $\tan\beta$--$M_{H^\pm}$ plane and the corresponding values for
the electron electric dipole moment are very similar to those displayed in
Figs.~\ref{075-110-112-100-300-200} and \ref{edm-077-110-112}.  For Set~B,
with ($M_1$, $M_2$, $\mu$)=(200, 400, 400)~GeV, the allowed region extends to
large values of $M_{H^\pm}$ without coming into conflict with the EW precision
data, and to large values of $\tan\beta$ without coming into conflict with
unitarity. However, the large-$\tan\beta$ part is constrained to
$M_{H^\pm}\sim400-500~\text{GeV}$ by the electroweak precision data, $T$ in
particular. This region roughly corresponds to the decoupling limit.  For
Set~C, there are two disconnected, allowed regions.

%%%%%%%%%%%%%%%%%%%%%%%%%%%%%%%%%%%%%%%%%%%%%%%%%%%%%%%%%%%%%%%%%%%%%%%%%%
\begin{figure}[htb]
%\vspace*{-2.0cm}
\centerline{
\includegraphics[width=7.5cm,angle=0]{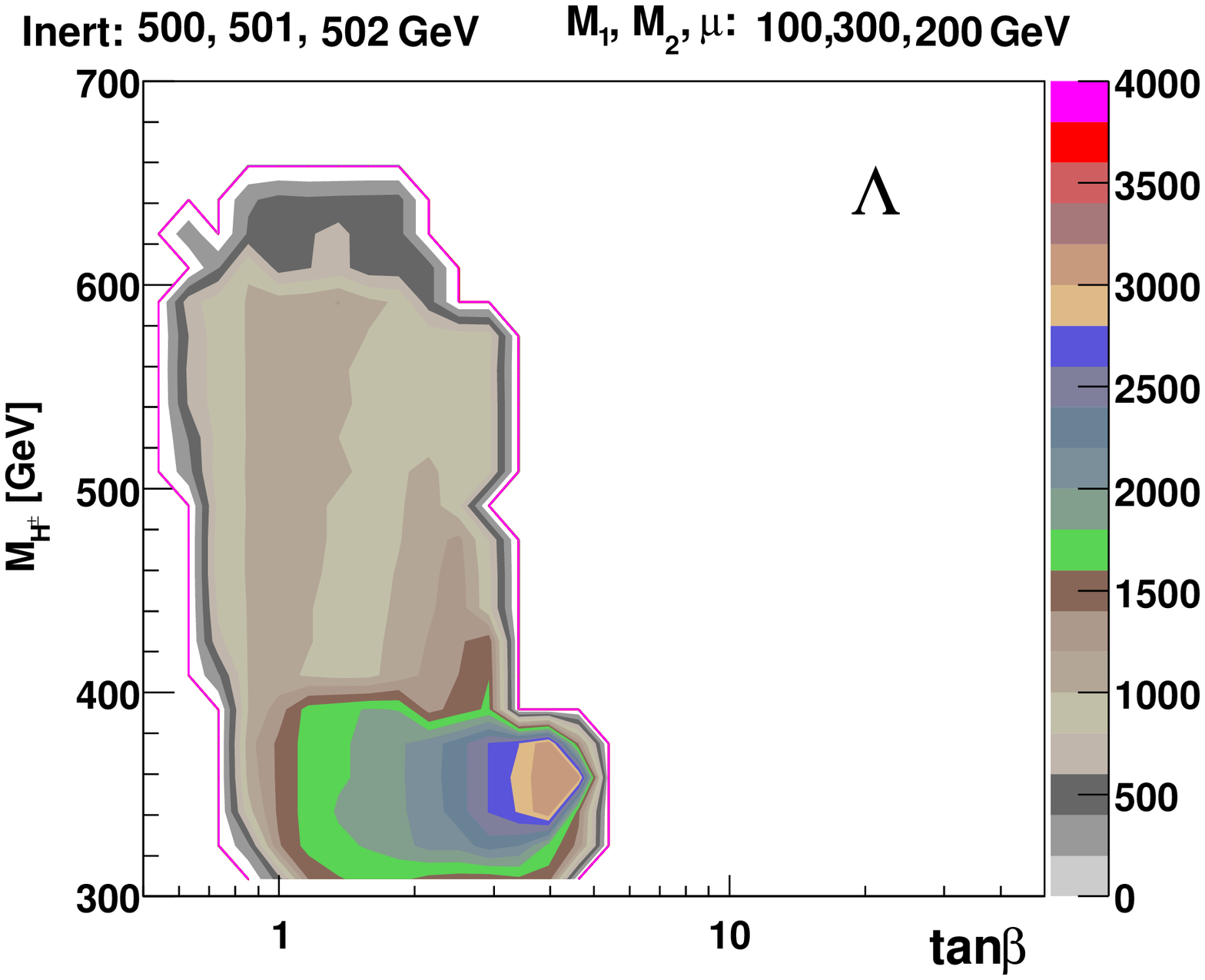}}
\centerline{
\includegraphics[width=7.5cm,angle=0]{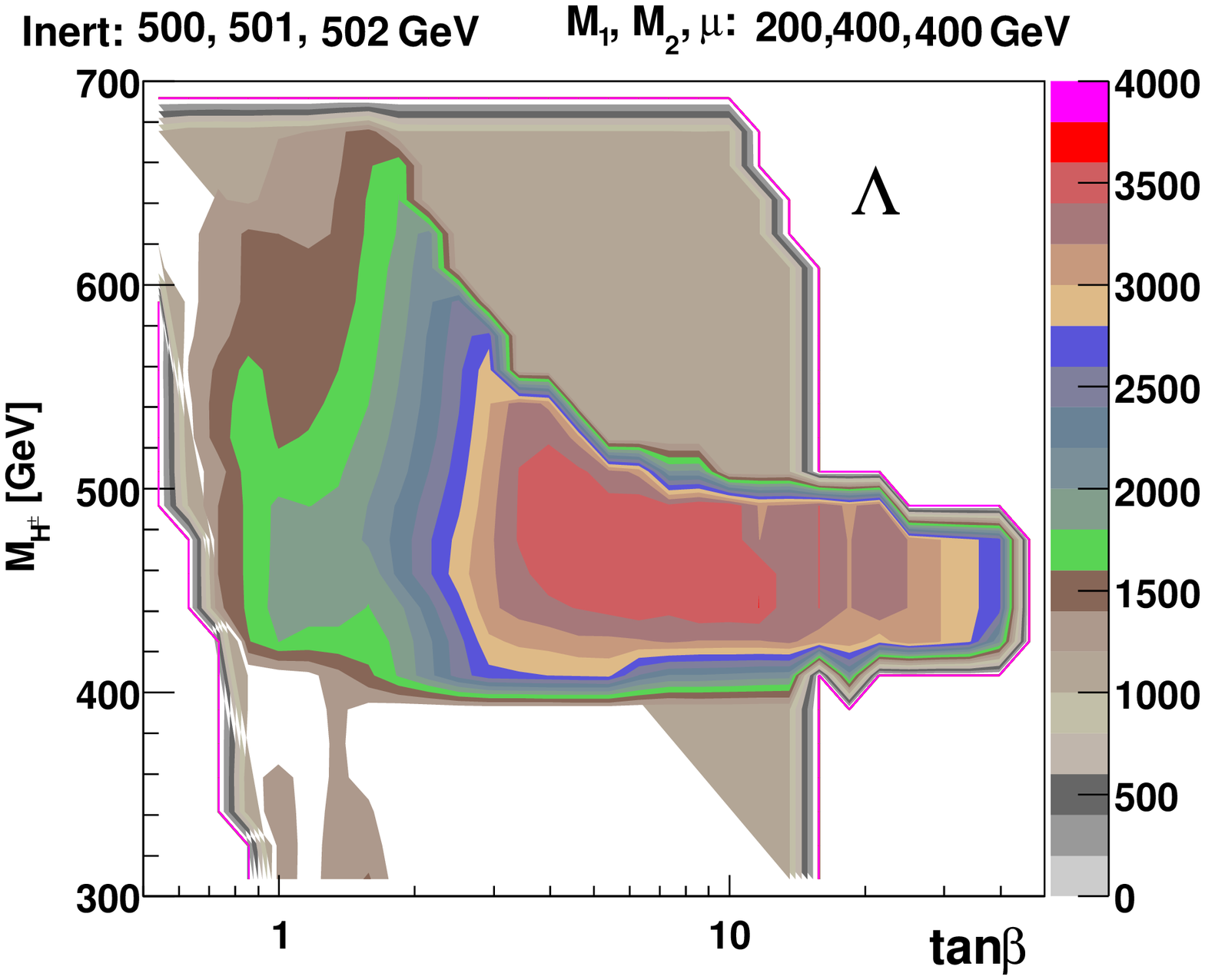}}
\centerline{
\includegraphics[width=7.5cm,angle=0]{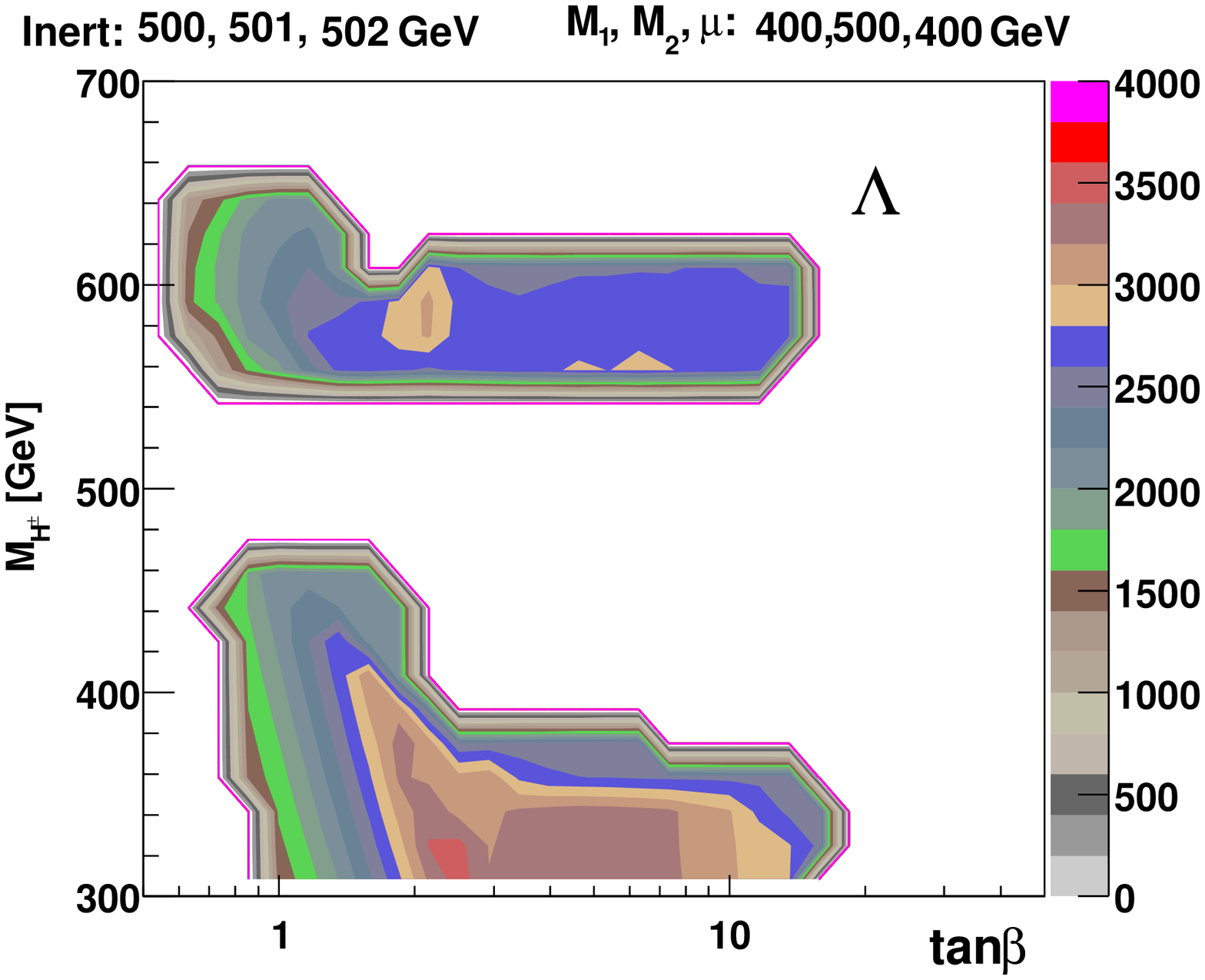}}
\caption{\label{cap_lam-500-501-502} Maximum reach in $\Lambda/\sqrt{D}$
[GeV]. Inert-sector masses: $(500, 501, 502)$~GeV; 2HDM-sector masses: ($M_1$,
$M_2$, $\mu$) = (100, 300, 200)~GeV,  (200, 400, 400)~GeV and 
(400, 500, 400)~GeV.}
\end{figure}
%%%%%%%%%%%%%%%%%%%%%%%%%%%%%%%%%%%%%%%%%%%%%%%%%%%%%%%%%%%%%%%%%%%%%%%%%%

{\bf The little hierarchy.}
We show in Fig.~\ref{cap_lam-500-501-502} the resulting ranges in
$\Lambda/\sqrt{D}$, as defined by Eq.~(\ref{Eq:cap_lambda}). This is seen to
reach well beyond 3~TeV, in particular for Set~B. Comparing sets of the 2HDM
mass parameters it is clear that, in agreement with our expectations, the
maximal value of the cut-off $\Lambda/\sqrt{D}$ grows with the mass scale of
the 2HDM model.  This illustrates our strategy to ameliorate the little
hierarachy problem by lifting the lowest visible Higgs mass. As already
mentioned, the allowed regions in this case are very similar to those
corresponding to light dark profiles shown in
Figs.~\ref{075-110-112-100-300-200} and \ref{edm-077-110-112}.  This is
consistent with the fact that the inert sector influences the experimental
constraints only through $T$ which is sensitive to inert scalar-mass
splitting. Since the splitting is of the same order (although it is larger for
the light dark profiles), therefore, it is not surprising that the allowed
regions are similar.  Consequently, if we impose a cut on
$\Lambda/\sqrt{D}=2~\text{TeV}$ we find allowed regions very similar to those
found for the lighter dark profiles, see Figs.~\ref{075-110-112-lam=2} and
\ref{077-110-112-lam=2}.

In general, the heavier non-inert Higgs states affect the amount of dark matter.
However, within the range of parameters explored, their effect can be compensated
by a retuning of the soft mass parameter $m_\eta$, which for fixed masses
yields a calibration of the trilinear inert--non-inert Higgs coupling $\lambda_L$.
%%%%%%%%%%%%%%%%%%%%%%%%%%%%%%%%%%%%%%%%%%%%%%%%%%%%%%%%%%%%%%%%%%%%%%%%%%%%%
\section{Summary}
\label{Sec:summary}
\setcounter{equation}{0}
%%%%%%%%%%%%%%%%%%%%%%%%%%%%%%%%%%%%%%%%%%%%%%%%%%%%%%%%%%%%%%%%%%%%%%%%%%%%%

We have explored an extension of the Inert Doublet
Model~\cite{Barbieri:2006dq,Deshpande:1977rw,Cao:2007rm} (IDM), made by
replacing the SM Higgs doublet sector by a 2-Higgs-Doublet Model (2HDM) in
order to accommodate CP violation in interactions of neutral Higgs bosons.
This model has four inert-sector scalars: two neutral, and a pair of charged
ones, as well as three ``ordinary'' neutral scalars and an accompanying pair
of charged ones. The latter five are those of the familiar 2HDM. Our
motivation here was not only to have CP violation in the scalar potential, but
also to provide a candidate for dark matter and to ameliorate the little
hierarchy problem by lifting Higgs boson masses (thereby increasing the
cut-off $\Lambda$).

We have estimated the amount of dark matter that is predicted by the model,
adopting the code {\tt micrOMEGAs}~\cite{Belanger:2006is,Belanger:2008sj},
checking all relevant theoretical and experimental constraints.  Solutions
were found both for ``light'' DM particles, with a mass around 75~GeV, as well
as for heavier ones, of the order of a few hundred GeV. In both cases, the
splitting between the masses of the charged and the neutral inert scalars must
be small in order to reproduce the right amount of dark matter.  In the
case of heavier dark matter the splitting is tiny implying nearly vanishing
contribution to the $T$ parameter.

As we have noted at the very end of Sec.~\ref{Subsec:Heavier_Dark
Matter_particle} the regions allowed by the condition $\Lambda/\sqrt{D} > 2
\tev$ for lighter and heavier dark matter profiles are similar. This is a
consequence of a very small contribution to $T$ from the inert
sector (because of the small mass splitting).  Since the inert sector
influences the experimental constraints only through $T$, the
allowed regions are similar. This observation illustrates an important
difference between this model and the original IDM, where the charged inert
scalar has to be considerably heavier than the neutral inert scalars in order
to provide a contribution to $T$ that can compensate a large negative SM
contribution (for a heavy SM Higgs boson). Here the role of the inert sector is
mainly restricted to providing a candidate for the dark matter, while $T$ can
be made consistent with the data utilizing only the freedom of the 2HDM
sector with negligible contribution from the inert sector. That freedom was
not available in the original inert model. In our approach the little
hierarchy problem is softened by increasing the Higgs boson masses,
this is possible mainly through the 2HDM sector alone.
 
In general, and in agreement with our expectations, the naturality arguments
favour heavier 2HDM masses. For instance, if one requires $\Lambda/\sqrt{D} >
2\tev$, then as seen in the top panel of Fig.~\ref{cap_lam-500-501-502}, only
a small region with $\tan\beta \sim 2-5$ and $M_{H^\pm} \sim 320-380\gev$ is
allowed for $(M_1,M_2,\mu)=(100,300,200)\gev$.  In contrast, for heavier 2HDM
masses (the middle panel in the same figure) a much larger region remains with
$\tan\beta \sim 1-40$ and $M_{H^\pm} \sim 400-640\gev$.  On the other hand,
for $(M_1,M_2,\mu)=(400,500,400)\gev$ (the lower panel) we observe again a
large allowed region, which however in this case is split into two
disconnected regions corresponding to light ($M_{H^\pm} \sim 300-450\gev$) and
heavy ($M_{H^\pm} \sim 550-650\gev$) charged Higgs bosons. The splitting is
related to partial cancellation between contributions to $T$ from charged and
neutral 2HDM scalars. It is worth noting, that the pictures which emerge
here are similar to the result obtained within the ordinary 2HDM
\cite{WahabElKaffas:2007xd} (see also \cite{Kanemura:2009mk}). As we have
already mentioned, that similarity follows from the small splitting in the
inert doublet-sector masses.

For the light-DM scenario discussed in Sec.~\ref{subsec:light-DM}, the
experimental implications (direct detection and production at the LHC) were
addressed in \cite{Barbieri:2006dq,Cao:2007rm,LopezHonorez:2006gr}. For the
case of heavier DM particles, the prospects are more dim, since the production
cross section will be rather small. For a detailed discussion, 
see \cite{Andreas:2009hj,Nezri:2009jd,Hambye:2009pw}.

As for the CP invariance, we observe that in general, CP can be substantially
violated in regions allowed by the experimental and theoretical
constraints. For instance, it is seen that the electron electric dipole moment
can easily exceed the allowed value of $10^{-27}[e\,\text{cm}]$ for
appropriate choices of the mixing angles.  We also note that it is easier to
accommodate CP violation at low values of $\tan\beta$ than at higher
values. For high values of $\tan\beta$, the allowed values of
$(\alpha_1,\alpha_2,\alpha_3)$ tend to accumulate near the limits where $H_2$
is odd under CP ($\alpha_2\simeq0$, $\alpha_3\simeq\pm\pi/2$).

It is also worth realizing that the model we discuss here bears some
similarity to the Weinberg model~\cite{Weinberg:1976hu} of CP violation with
natural absence of flavour-changing Yukawa couplings.  Both models invoke
three doublets, one of which has no Yukawa couplings while the two
others couple to fermions such that no flavour-changing Yukawa couplings
emerge.  However, the important difference is that here (in order to
guarantee stability of the dark matter candidate) we impose an extra $Z_2$
symmetry ($\eta\to -\eta$) which remains unbroken after spontaneous symmetry
breaking since $\langle \eta \rangle =0$.  As a consequence, there is
no mixing in the mass matrix between the charged components of $\Phi_{1,2}$ and
$\eta$, therefore eventually $\eta^\pm$ is a mass eigenstate with no Yukawa
couplings at all. Then CP violation in the charged scalar sector is the same
as in a pure 2HDM with Yukawa couplings parameterized by the CKM matrix alone.

It should be noted that restricting our study to the range of inert-model
parameters satisfying (\ref{posit}), rather than utilizing the full range
defined by Eqs.~(\ref{constraints}) and (\ref{Eq:positivity-cond.2}), we are
clearly not able to find all allowed domains in the parameter space.  However,
a full investigation is technically much more involved and therefore
computationally more challenging.  In order to determine all the quartic
couplings in the 2DHM sector ($V_{12}$) one has to specify $(M_1,M_2,\mu)$,
$\tan\beta$, $M_{H^\pm}$ and also the angles $(\alpha_1,\alpha_2,\alpha_3)$.
To fix the quartic couplings between the 2HDM and the inert sector contained
in $V_{123}$ one has to know also the inert mass parameters
$(M_S,M_A,M_{\eta^\pm})$ and $m_\eta$.  Then for each point in the parameter
space (including angles over which we scan) both Eqs.~(\ref{constraints}) and
(\ref{Eq:positivity-cond.2}) must be checked. In the presence of the large
number of free parameters, that makes the analysis much more complicated and
time consuming.  Also, within the general strategy outlined here, the
calculation of the DM abundance would be much more complicated, since for that
purpose all the parameters must be simultaneously known. That is indeed
necessary as cross sections for dark matter annihilation depend on the mixing
angles $(\alpha_1,\alpha_2,\alpha_3)$, masses of the 2HDM scalars
$M_1,M_2,M_3$ and, of course, also on the inert mass parameters
$(M_S,M_A,M_{\eta^\pm})$ and $m_\eta$.  A more complete investigation adopting
the general positivity conditions Eqs.~(\ref{constraints}) and
(\ref{Eq:positivity-cond.2}) together with a more precise calculation of the
dark matter abundance will be attempted, and reported on elsewhere.

\vspace*{10mm}
{\bf Acknowledgements.}  We thank the NORDITA program {\it ``TeV scale physics
and dark matter''}, for hospitality while this work was initiated. It is a
pleasure to thank M.~Gruenewald for advice on the LEP1 electroweak precision
data.  We are also grateful to U.~Nierste, S.~Trine and S.~Westhoff for advice
on the $B\to D\tau\bar\nu_\tau$ analysis, as well as to G.~Belanger,
A.~Pukhov and M.~Tytgat for help with the installation of {\tt micrOMEGAs} and
running the IDM. We thank J.~Gunion for his interest at the beginning of this
project.  The research of P.O. has been supported by the Research Council of
Norway. This research of B.G. is supported in part by the Ministry of Science
and Higher Education (Poland) as research project N~N202~006334 (2008-11).
B.G. also acknowledges support of the European Community within the Marie
Curie Research \& Training Networks: ``HEPTOOLS" (MRTN-CT-2006-035505), and
``UniverseNet" (MRTN-CT-2006-035863), and through the Marie Curie Host
Fellowships for the Transfer of Knowledge Project MTKD-CT-2005-029466.

%%%%%%%%%%%%%%%%%%%%%%%%%%%%%%%%%%%%%%%%%%%%%%%%%%%%%%%%%%%%%%%%%%%%%%
\appendix
%%%%%%%%%%%%%%%%%%%%%%%%%%%%%%%%%%%%%%%%%%%%%%%%%%%%%%%%%%%%%%%%%%%%%%

%%%%%%%%%%%%%%%%%%%%%%%%%%%%%%%%%%%%%%%%%%%%%%%%%%%%%%%%%%%%%%%%%%%%%%%%%%%%%
\section{Positivity of the potential}
\setcounter{equation}{0}
%\renewcommand{\thesection}{A}
%%%%%%%%%%%%%%%%%%%%%%%%%%%%%%%%%%%%%%%%%%%%%%%%%%%%%%%%%%%%%%%%%%%%%%%%%%%%%

In order to ensure vacuum stability, the potential should be positive for large
values of the fields, $\Phi_1$, $\Phi_2$ and $\eta$.
We study the general potential
\begin{eqnarray}
V(\Phi_1,\Phi_2,\eta) 
&=&\frac{\lambda_1}{2}(\Phi_1^\dagger\Phi_1)^2 
+ \frac{\lambda_2}{2}(\Phi_2^\dagger\Phi_2)^2
+ \frac{\lambda_\eta}{2} (\eta^\dagger \eta)^2 \nonumber \\
&&+ \lambda_3(\Phi_1^\dagger\Phi_1)(\Phi_2^\dagger\Phi_2) 
+ \lambda_4(\Phi_1^\dagger\Phi_2)(\Phi_2^\dagger\Phi_1)
+ \frac12\left[\lambda_5(\Phi_1^\dagger\Phi_2)^2 + \hc\right] \nonumber\\
&&+\lambda_{1133} (\Phi_1^\dagger\Phi_1)(\eta^\dagger \eta)
+\lambda_{2233} (\Phi_2^\dagger\Phi_2)(\eta^\dagger \eta)\nonumber\\
&&+\lambda_{1331}(\Phi_1^\dagger\eta)(\eta^\dagger\Phi_1) 
+\lambda_{2332}(\Phi_2^\dagger\eta)(\eta^\dagger\Phi_2) \nonumber  \\
&& 
+\half\left[\lambda_{1313}(\Phi_1^\dagger\eta)^2 +\hc \right]  
+\half\left[\lambda_{2323}(\Phi_2^\dagger\eta)^2 +\hc \right] \nonumber\\
&&-\frac12\left\{m_{11}^2\Phi_1^\dagger\Phi_1 
+ m_{22}^2\Phi_2^\dagger\Phi_2 + \left[m_{12}^2 \Phi_1^\dagger \Phi_2 
+ \hc\right]\right\}\nonumber\\
&&+m_\eta^2\eta^\dagger \eta.
\end{eqnarray}
We start by rewriting the Higgs doublets as:
\begin{align}
\label{Eq:normphi_i}
\Phi_1=||\Phi_1||{\hat{\Phi}}_1,\hspace*{1cm}
\Phi_2=||\Phi_2||{\hat{\Phi}}_2,\hspace*{1cm}
\eta = ||\eta||{\hat{\eta}},
\end{align}
where $||\Phi_i||$ and $||\eta||$ are the norms of the spinors, 
and ${\hat{\Phi}}_i$ and ${\hat{\eta}}$ are unit spinors. By $SU(2)$
invariance, only the following combinations of fields may appear:
\begin{eqnarray}
&&\Phi_1^\dagger\Phi_1=||\Phi_1||^2, \quad 
\Phi_2^\dagger\Phi_2=||\Phi_2||^2, \quad
\eta^\dagger\eta=||\eta||^2, \nonumber\\
&&\Phi_2^\dagger\Phi_1=||\Phi_1||\cdot||\Phi_2||
\left({\hat{\Phi}}_2^\dagger\cdot{\hat{\Phi}}_1\right), \quad
\Phi_1^\dagger\Phi_2=[\Phi_2^\dagger\Phi_1]^*,\nonumber\\
&&\eta^\dagger\Phi_1=||\Phi_1||\cdot||\eta||
\left({\hat{\eta}}^\dagger\cdot{\hat{\Phi}}_1\right), \quad
\Phi_1^\dagger\eta=[\eta^\dagger\Phi_1]^*,\nonumber\\
&&\eta^\dagger\Phi_2=||\Phi_2||\cdot||\eta||
\left({\hat{\eta}}^\dagger\cdot{\hat{\Phi}}_2\right), \quad
\Phi_2^\dagger\eta=[\eta^\dagger\Phi_2]^*.
\end{eqnarray}
We let the norms of Eq.~(\ref{Eq:normphi_i}) be parametrized as follows:
\begin{equation}
\label{Eq:parametrization1}
||\Phi_1||=r\cos\gamma\sin\theta, \qquad
||\Phi_2||=r\sin\gamma\sin\theta, \qquad
||\eta||=r\cos\theta.
\end{equation}
The complex product between two different unit spinors will be a 
complex number with modulus less than or equal to unity, i.e.
\begin{equation}
\label{Eq:parametrization2}
{\hat{\Phi}}_2^\dagger\cdot{\hat{\Phi}}_1=\rho_1e^{i\phi_1}, \qquad
{\hat{\eta}}^\dagger\cdot{\hat{\Phi}}_1=\rho_2e^{i\phi_2}, \qquad
{\hat{\eta}}^\dagger\cdot{\hat{\Phi}}_2=\rho_3e^{i\phi_3}.
\end{equation}
Using this parametrization, we can write:
\begin{align}
\Phi_1^\dagger\Phi_1&=r^2\cos^2\gamma\sin^2\theta, \quad 
\Phi_2^\dagger\Phi_2=r^2\sin^2\gamma\sin^2\theta, \quad
\eta^\dagger\eta=r^2\cos^2\theta,\nonumber \\
\Phi_2^\dagger\Phi_1&=r^2\cos\gamma\sin\gamma\sin^2\theta
\rho_1e^{i\phi_1}, \quad
\Phi_1^\dagger\Phi_2=r^2\cos\gamma\sin\gamma\sin^2\theta\rho_1e^{-i\phi_1},
\nonumber\\
\eta^\dagger\Phi_1&=r^2\cos\gamma\sin\theta\cos\theta
\rho_2e^{i\phi_2}, \quad
\Phi_1^\dagger\eta=r^2\cos\gamma\sin\theta\cos\theta\rho_2e^{-i\phi_2},
\nonumber\\
\eta^\dagger\Phi_2&=r^2\sin\gamma\sin\theta\cos\theta
\rho_3e^{i\phi_3}, \quad
\Phi_2^\dagger\eta=r^2\sin\gamma\sin\theta\cos\theta\rho_3e^{-i\phi_3}
\end{align}
where $r\geq0$, $\gamma\in[0,\pi/2]$, $\theta\in[0,\pi/2]$,
$\rho_i\in[0,1]$ and $\phi_i\in[0,2\pi\rangle$.

The potential can now be written as
\begin{equation}
V=r^4V_4+r^2V_2,
\end{equation}
with only the quartic, $V_4$, part relevant for positivity:
\begin{eqnarray}
\label{Eq:pot_4}
V_4&=&\lambda_1A_1+\lambda_2A_2+\lambda_\eta A_3+\lambda_3A_4+\lambda_4A_5
\nonumber\\
&&+\lambda_{1133}A_6+\lambda_{2233}A_7+\lambda_{1331}A_8+\lambda_{2332}A_9
\nonumber\\
&&+\Re\lambda_5A_{10}+\Im\lambda_5A_{11}\nonumber\\
&&+\Re\lambda_{1313}A_{12}+\Im\lambda_{1313}A_{13}
+\Re\lambda_{2323}A_{14}+\Im\lambda_{2323}A_{15},
\end{eqnarray}
where
\begin{align}
\label{Eq:A_i2}
A_1&=\frac{1}{2}\cos^4\gamma\sin^4\theta, \quad
A_2=\frac{1}{2}\sin^4\gamma\sin^4\theta, \quad
A_3=\frac{1}{2}\cos^4\theta, \quad
A_4=\cos^2\gamma\sin^2\gamma\sin^4\theta, \nonumber\\
A_5&=\rho_1^2\cos^2\gamma\sin^2\gamma\sin^4\theta,\quad
A_6=\cos^2\gamma\sin^2\theta\cos^2\theta,\quad
A_7=\sin^2\gamma\sin^2\theta\cos^2\theta, \nonumber\\
A_8&=\rho_2^2\cos^2\gamma\sin^2\theta\cos^2\theta, \quad
A_9=\rho_3^2\sin^2\gamma\sin^2\theta\cos^2\theta, \nonumber\\
A_{10}&=\rho_1^2\cos(2\phi_1)\cos^2\gamma\sin^2\gamma\sin^4\theta,\quad
A_{11}=\rho_1^2\sin(2\phi_1)\cos^2\gamma\sin^2\gamma\sin^4\theta,\nonumber \\
A_{12}&=\rho_2^2\cos(2\phi_2)\cos^2\gamma\sin^2\theta\cos^2\theta, \quad
A_{13}=\rho_2^2\sin(2\phi_2)\cos^2\gamma\sin^2\theta\cos^2\theta,\nonumber \\
A_{14}&=\rho_3^2\cos(2\phi_3)\sin^2\gamma\sin^2\theta\cos^2\theta,\quad
A_{15}=\rho_3^2\sin(2\phi_3)\sin^2\gamma\sin^2\theta\cos^2\theta.
\end{align}

The quartic part of the potential can now be written:
\begin{eqnarray}
V_4&=&\frac{\lambda_1}{2}\cos^4\gamma\sin^4\theta
+\frac{\lambda_2}{2}\sin^4\gamma\sin^4\theta
+\frac{\lambda_\eta}{2}\cos^4\theta\nonumber\\
&&+\lambda_3\cos^2\gamma\sin^2\gamma\sin^4\theta
+\lambda_{1133}\cos^2\gamma\sin^2\theta\cos^2\theta
+\lambda_{2233}\sin^2\gamma\sin^2\theta\cos^2\theta\nonumber\\
&&+\rho_1^2\left[\lambda_4+\Re\lambda_5\cos(2\phi_1)
+\Im\lambda_5\sin(2\phi_1)\right]\cos^2\gamma\sin^2\gamma\sin^4\theta
\nonumber\\
&&+\rho_2^2\left[\lambda_{1331}+\Re\lambda_{1313}\cos(2\phi_2)
+\Im\lambda_{1313}\sin(2\phi_2)\right]\cos^2\gamma\sin^2\theta\cos^2\theta
\nonumber\\
&&+\rho_3^2\left[\lambda_{2332}+\Re\lambda_{2323}\cos(2\phi_3)
+\Im\lambda_{2323}\sin(2\phi_3)\right]\sin^2\gamma\sin^2\theta\cos^2\theta.
\end{eqnarray}
We minimize this expression with respect to $\phi_i$ to arrive at
\begin{eqnarray}
\bar{V}_4&&\frac{\lambda_1}{2}\cos^4\gamma\sin^4\theta
+\frac{\lambda_2}{2}\sin^4\gamma\sin^4\theta
+\frac{\lambda_\eta}{2}\cos^4\theta\nonumber\\
&&+\lambda_3\cos^2\gamma\sin^2\gamma\sin^4\theta
+\lambda_{1133}\cos^2\gamma\sin^2\theta\cos^2\theta
+\lambda_{2233}\sin^2\gamma\sin^2\theta\cos^2\theta\nonumber\\
&&+\rho_1^2\left(\lambda_4-|\lambda_5|\right)
\cos^2\gamma\sin^2\gamma\sin^4\theta\nonumber\\
&&+\rho_2^2\left(\lambda_{1331}-|\lambda_{1313}|\right)
\cos^2\gamma\sin^2\theta\cos^2\theta\nonumber\\
&&+\rho_3^2\left(\lambda_{2332}-|\lambda_{2323}|\right)
\sin^2\gamma\sin^2\theta\cos^2\theta.
\end{eqnarray}
Further, we minimize this expression with respect to $\rho_i$ to arrive at:
\begin{eqnarray}
\tilde{V}_4&&\frac{\lambda_1}{2}\cos^4\gamma\sin^4\theta
+\frac{\lambda_2}{2}\sin^4\gamma\sin^4\theta
+\frac{\lambda_\eta}{2}\cos^4\theta\nonumber\\
&&+\lambda_x\cos^2\gamma\sin^2\gamma\sin^4\theta
+\lambda_y\cos^2\gamma\sin^2\theta\cos^2\theta
+\lambda_z\sin^2\gamma\sin^2\theta\cos^2\theta.
\end{eqnarray}
where
\begin{eqnarray}
\lambda_x&=&\lambda_3+\min\left(0,\lambda_4-|\lambda_5|\right)\\
\lambda_y&=&\lambda_{1133}+\min\left(0,\lambda_{1331}-|\lambda_{1313}|\right)\\
\lambda_z&=&\lambda_{2233}+\min\left(0,\lambda_{2332}-|\lambda_{2323}|\right)
\end{eqnarray}
For the positivity condition to be satisfied, $\tilde{V}_4$ must be positive
for all combinations of $\gamma\in[0,\pi/2]$ and $\theta\in[0,\pi/2]$.  This
is both a necessary and a sufficient condition.  

%%%%%%%%%%%%%%%%%%%%%%%%%%%%%%%%%%%%%%%%%%%%%%%%%%%%%%%%%%%%%%%%%%%%%%%%%%%%%
\subsection{Boundary points}
%%%%%%%%%%%%%%%%%%%%%%%%%%%%%%%%%%%%%%%%%%%%%%%%%%%%%%%%%%%%%%%%%%%%%%%%%%%%%
Some points from the
parameter space give us some rather simple positivity conditions. We now turn
our attention towards these special points.\\ \\
\noindent{\underline{\bf $\theta=0$ or $\theta=\pi/2$ or $\gamma=0$ 
or $\gamma=\pi/2$}}\\
\\
First we consider the boundary points in the $(\gamma,\theta)$ plane.
\begin{eqnarray}
\tilde{V}_4(\theta=0)&=&\frac{\lambda_\eta}{2},\nonumber\\
\tilde{V}_4(\theta=\frac{\pi}{2})&=&\frac{\lambda_1}{2}\cos^4\gamma
+\frac{\lambda_2}{2}\sin^4\gamma+\lambda_x\cos^2\gamma\sin^2\gamma,\nonumber\\
\tilde{V}_4(\gamma=0)&=&\frac{\lambda_1}{2}\sin^4\theta
+\frac{\lambda_\eta}{2}\cos^4\theta+\lambda_y\sin^2\theta\cos^2\theta,
\nonumber\\
\tilde{V}_4(\gamma=\frac{\pi}{2})&=&\frac{\lambda_2}{2}\sin^4\theta
+\frac{\lambda_\eta}{2}\cos^4\theta+\lambda_z\sin^2\theta\cos^2\theta.\nonumber
\end{eqnarray}
The last three of these expressions have the same form as an expression
already studied in the 2HDM \cite{El Kaffas:2006nt}. 
Using a result from there, we end up with the
following conditions:
\begin{equation} 
\lambda_1>0,\quad \lambda_2>0,\quad \lambda_\eta>0,\quad 
\lambda_x>-\sqrt{\lambda_1\lambda_2},
\quad \lambda_y>-\sqrt{\lambda_1\lambda_\eta},\quad 
\lambda_z>-\sqrt{\lambda_2\lambda_\eta}\label{constraints}.
\end{equation}
%%%%%%%%%%%%%%%%%%%%%%%%%%%%%%%%%%%%%%%%%%%%%%%%%%%%%%%%%%%%%%%%%%%%%%%%%%%%%
\subsection{Interior points}
%%%%%%%%%%%%%%%%%%%%%%%%%%%%%%%%%%%%%%%%%%%%%%%%%%%%%%%%%%%%%%%%%%%%%%%%%%%%%
What remains is to demand that $\tilde{V}_4 >0$ also in the interior of
the $(\gamma,\theta)$ plane. Thus,
\begin{eqnarray}
\lambda_y\cos^2\gamma+\lambda_z\sin^2\gamma>
-\frac{1}{2}\left\{\frac{\lambda_\eta}{\tan^2\theta}+
(\lambda_1\cos^4\gamma+\lambda_2\sin^4\gamma
+2\lambda_x\cos^2\gamma\sin^2\gamma)\tan^2\theta\right\}\nonumber
\end{eqnarray}
Maximizing the right hand side of this inequality with respect to $\theta$, we
find that the maximum occurs at
$\tan^2\theta=\sqrt{\lambda_\eta/(\lambda_1\cos^4\gamma
+\lambda_2\sin^4\gamma+2\lambda_x\cos^2\gamma\sin^2\gamma)}$.
Substituting this back we arrive at:
\begin{eqnarray}
\lambda_y\cos^2\gamma+\lambda_z\sin^2\gamma
>-\sqrt{\lambda_\eta(\lambda_1\cos^4\gamma+\lambda_2\sin^4\gamma
+2\lambda_x\cos^2\gamma\sin^2\gamma)}\label{finalineq}
\end{eqnarray}
We need to solve this inequality subject to the constraints given in
(\ref{constraints}).  Let us distinguish between four different cases:\\ \\
\noindent{\bf \underline{Case a) $\lambda_y\geq0$ and $\lambda_z\geq0$}}\\ \\
The inequality (\ref{finalineq}) is trivially satisfied.\\ \\
\noindent{\bf \underline{Case b) $\lambda_y>0$ and $\lambda_z<0$}}\\ \\
The left-hand side of (\ref{finalineq}) can be both positive and negative. 
See details in section~\ref{Sect:cases_b_c}. \\ \\
\noindent{\underline{\bf Case c) $\lambda_y<0$ and $\lambda_z>0$}}\\ \\
The left-hand side of (\ref{finalineq}) can be both positive and negative. 
See details in section~\ref{Sect:cases_b_c}.
Should be similar to case b) with $\lambda_y$ and $\lambda_z$ interchanged.
\\ \\
\noindent{\underline{\bf Case d) $(\lambda_y\leq0 \wedge \lambda_z<0)$ or
$(\lambda_y<0 \wedge \lambda_z\leq0)$}}\\ \\ We can now square both sides of
the inequality and reverse the inequality sign to get
\begin{eqnarray}
\lambda_y^2\cos^4\gamma+2\lambda_y\lambda_z\cos^2\gamma\sin^2\gamma
+\lambda_z^2\sin^4\gamma
<\lambda_\eta(\lambda_1\cos^4\gamma+\lambda_2\sin^4\gamma
+2\lambda_x\cos^2\gamma\sin^2\gamma)\nonumber
\end{eqnarray}
and finally
\begin{eqnarray}
(\lambda_\eta\lambda_1-\lambda_y^2)\cos^4\gamma
+(\lambda_\eta\lambda_2-\lambda_z^2)\sin^4\gamma
+2(\lambda_\eta\lambda_x-\lambda_y\lambda_z)\cos^2\gamma\sin^2\gamma>0\nonumber
\end{eqnarray}
which is positive definite if 
\begin{eqnarray}
\lambda_\eta\lambda_x-\lambda_y\lambda_z>
-\sqrt{(\lambda_\eta\lambda_1-\lambda_y^2)(\lambda_\eta\lambda_2-\lambda_z^2)}
\end{eqnarray}
%%%%%%%%%%%%%%%%%%%%%%%%%%%%%%%%%%%%%%%%%%%%%%%%%%%%%%%%%%%%%%%%%%%%%%%%%%%%%
\subsection{A detailed study of positivity in cases b) and c)}
\label{Sect:cases_b_c}
%%%%%%%%%%%%%%%%%%%%%%%%%%%%%%%%%%%%%%%%%%%%%%%%%%%%%%%%%%%%%%%%%%%%%%%%%%%%%
We consider the inequality (\ref{finalineq}):
\begin{eqnarray}
\lambda_y\cos^2\gamma+\lambda_z\sin^2\gamma
>-\sqrt{\lambda_\eta(\lambda_1\cos^4\gamma+\lambda_2\sin^4\gamma
+2\lambda_x\cos^2\gamma\sin^2\gamma)}.
\end{eqnarray}
Let us introduce $x=\tan^2\gamma$ and transform the inequality into
\begin{eqnarray}
\lambda_y+\lambda_z x
>-\sqrt{\lambda_\eta(\lambda_1+\lambda_2x^2+2\lambda_x x)}.\label{inequalityx}
\end{eqnarray}
which must be satisfied for $x>0$.
In order to analyze this we will study the solutions of the equation
\begin{eqnarray}
\lambda_y+\lambda_z x
=-\sqrt{\lambda_\eta(\lambda_1+\lambda_2x^2+2\lambda_x x)}.\label{realequation}
\end{eqnarray}
Possible solutions of this equation are given by
\begin{eqnarray}
x_{1,2}=\frac{-b\pm\sqrt{b^2-4ac}}{2a}
\end{eqnarray}
where $a=\lambda_\eta\lambda_2-\lambda_z^2$, $b=2(\lambda_\eta\lambda_x
-\lambda_y\lambda_z)$ and $c=\lambda_\eta\lambda_1-\lambda_y^2$.
These possible solutions are obtained by squaring (\ref{realequation}). 
However, in doing this we may introduce false solutions, that is solutions of
\begin{eqnarray}
\lambda_y+\lambda_z x
=+\sqrt{\lambda_\eta(\lambda_1+\lambda_2x^2+2\lambda_x x)}.
\label{falseequation}
\end{eqnarray}
In order for (\ref{inequalityx}) to be satisfied we must demand that
(\ref{realequation}) does not have any positive real-valued solutions when
subject to the constraints already obtained in (\ref{constraints}).  That is:
\begin{eqnarray}
\overline{
b^2-4ac\geq0 \wedge\left[
\left(x_1>0 \wedge \lambda_y+\lambda_z x_1<0\right) \vee
\left(x_2>0 \wedge \lambda_y+\lambda_z x_2<0\right) \right]}
\end{eqnarray}
where the overline (bar) denotes negation.
This expression must be true in order for positivity to be satisfied. 
The possible solutions read:\\ \\
\noindent{\bf \underline{Case b) $\lambda_y>0$ and $\lambda_z<0$}}\\ \\ In
this case $a>0$, while $c$ can be both positive, zero or negative. We have to
distinguish between these cases. The results are:
\begin{eqnarray}
\left[c>0 \wedge \left(a\geq \frac{c\lambda_z^2}{\lambda_y^2} 
\vee b>-2\sqrt{ac}\right)\right]\vee c\leq0
\end{eqnarray}
equivalent to
\begin{eqnarray}
\left[\lambda_\eta\lambda_1>\lambda_y^2 \wedge \left(\lambda_2\lambda_y^2
\geq\lambda_1\lambda_z^2 \vee 
\lambda_\eta\lambda_x-\lambda_y\lambda_z>
-\sqrt{(\lambda_\eta\lambda_1-\lambda_y^2)
(\lambda_\eta\lambda_2-\lambda_z^2)}\right)\right] 
\vee \lambda_\eta\lambda_1\leq\lambda_y^2\nonumber\\
\end{eqnarray}
\noindent{\bf \underline{Case c) $\lambda_y<0$ and $\lambda_z>0$}}\\ \\
In this case $c>0$, while $a$ can be both positive, zero or negative. 
We need to distinguish between these cases. The results are:
\begin{eqnarray}
\left[a>0 \wedge \left(c\geq \frac{a\lambda_y^2}{\lambda_z^2} 
\vee b>-2\sqrt{ac}\right)\right]\vee a\leq0
\end{eqnarray}
equivalent to
\begin{eqnarray}
\left[\lambda_\eta\lambda_2>\lambda_z^2 \wedge \left(\lambda_1\lambda_z^2
\geq\lambda_2\lambda_y^2 \vee 
\lambda_\eta\lambda_x-\lambda_y\lambda_z>
-\sqrt{(\lambda_\eta\lambda_1-\lambda_y^2)
(\lambda_\eta\lambda_2-\lambda_z^2)}\right)\right] 
\vee \lambda_\eta\lambda_2\leq\lambda_z^2\nonumber\\
\end{eqnarray}
Combining the results from cases a) - d) subject to the constraints of
(\ref{constraints}) there are remarkable(!) simplifications. We end up with
\begin{equation}
\sqrt{\lambda_1}\lambda_z+\sqrt{\lambda_2}\lambda_y\geq0 \quad \vee \quad
\lambda_\eta\lambda_x-\lambda_y\lambda_z>
-\sqrt{(\lambda_\eta\lambda_1-\lambda_y^2)(\lambda_\eta\lambda_2-\lambda_z^2)}
\end{equation}
as an additional constraint to the ones listed in (\ref{constraints}),
or expressed in a more symmetric form:
\begin{equation}
\sqrt{\lambda_\eta}\lambda_x+\sqrt{\lambda_1}\lambda_z
+\sqrt{\lambda_2}\lambda_y\geq0 \quad \vee \quad
\lambda_\eta\lambda_x^2+\lambda_1\lambda_z^2+\lambda_2\lambda_y^2
-\lambda_\eta\lambda_1\lambda_2-2\lambda_x\lambda_y\lambda_z<0
\label{finalconstraints}
\end{equation}

%%%%%%%%%%%%%%%%%%%%%%%%%%%%%%%%%%%%%%%%%%%%%%%%%%%%%%%%%%%%%%%%%%%%%%%%%%%%%
\subsection{The dark democracy}
%%%%%%%%%%%%%%%%%%%%%%%%%%%%%%%%%%%%%%%%%%%%%%%%%%%%%%%%%%%%%%%%%%%%%%%%%%%%%
In the so-called dark democracy (\ref{Eq:DarkDemocracy}),
$\lambda_y=\lambda_z$ and (\ref{finalineq}) simplifies to
\begin{eqnarray}
\lambda_y
>-\sqrt{\lambda_\eta(\lambda_1\cos^4\gamma+\lambda_2\sin^4\gamma
+2\lambda_x\cos^2\gamma\sin^2\gamma)}
\end{eqnarray}
which is satisfied whenever 
\begin{eqnarray} \label{Eq:positivity-cond.2}
\lambda_y\geq0 \vee \left(\lambda_\eta\lambda_x-\lambda_y^2
>-\sqrt{(\lambda_\eta\lambda_1-\lambda_y^2)
(\lambda_\eta\lambda_2-\lambda_y^2)}\right)
\end{eqnarray}
This constraint combined with the inequalities from (\ref{constraints}) forms
a necessary and sufficient condition to guarantee positivity of the potential.

%%%%%%%%%%%%%%%%%%%%%%%%%%%%%%%%%%%%%%%%%%%%%%%%%%%%%%%%%%%%%%%%%%%%%%%%%%%%%
\section{CP conservation}
\setcounter{equation}{0}
\setcounter{subsection}{0}
%\renewcommand{\thesection}{C}
%%%%%%%%%%%%%%%%%%%%%%%%%%%%%%%%%%%%%%%%%%%%%%%%%%%%%%%%%%%%%%%%%%%%%%%%%%%%%

In the general case, the mixing of the different weak neutral states is
described by the angles $\{\alpha_1,\alpha_2,\alpha_3\}$ present in the
rotation matrix $R$. There are three simple limits of no CP violation, namely
when either $H_1$, $H_2$ or $H_3$ is odd under CP. These limits can all be
defined in terms of $\alpha_2$ and $\alpha_3$, as illustrated in Fig.~1 of
\cite{ElKaffas:2007rq}: $H_1$ is odd when $\alpha_2=\pm\pi/2$, $H_2$ is odd
when $\alpha_2=0$ and $\alpha_3=\pm\pi/2$, whereas $H_3$ is odd when
$\alpha_2=\alpha_3=0$.  Away from these limits, the model in general violates
CP.  (Exceptions will be discussed below.)

\begin{figure}[htb]
%\vspace*{-2.0cm}
\centerline{
\includegraphics[width=15.5cm,angle=0]{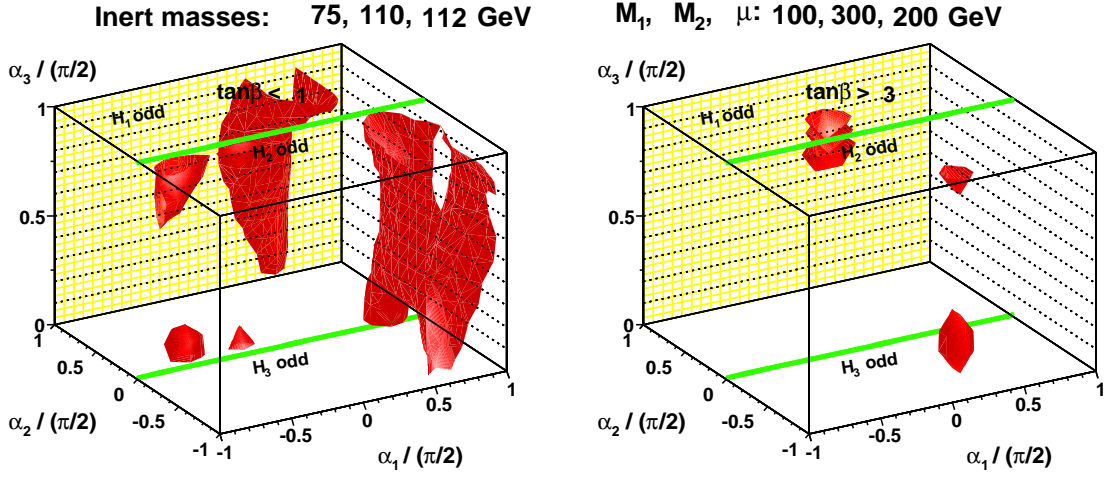}
}
\caption{\label{alphas-075-110-112-100-300-200} Populated regions of
$\alpha$-space. Left: $\tan\beta<1$; Right: $\tan\beta>3$.
Inert-sector masses: $(75, 110, 112)$~GeV; 2HDM-sector masses:
($M_1$, $M_2$, $\mu$) = (100, 300, 200)~GeV. Green lines at $\alpha_2=0$
(and $\alpha_3=\pi/2$ or 0)
show limits of no CP violation, with either $H_2$ or $H_3$ being odd under CP.
The limit $\alpha_2=\pi/2$ (yellow plane) indicates where $H_1$ is odd. There is a corresponding one
at $\alpha_2=-\pi/2$ (not shown).}
\end{figure}

\begin{figure}[htb]
%\vspace*{-2.0cm}
\centerline{
\includegraphics[width=15.5cm,angle=0]{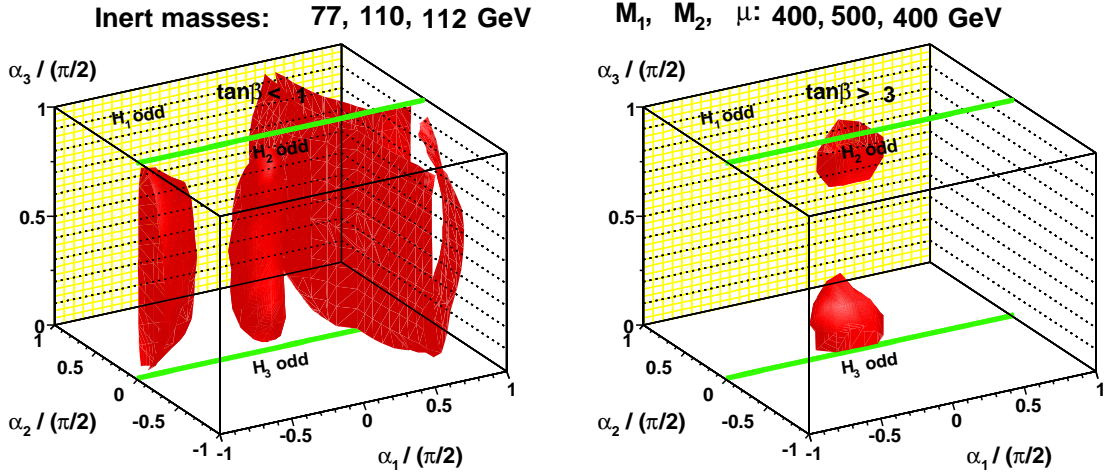}
}
\caption{\label{alphas-077-110-112-400-500-400} Populated regions of
$\alpha$-space. Similar to Fig.~\ref{alphas-075-110-112-100-300-200} for
 inert-sector masses: $(77, 110, 112)$~GeV; 2HDM-sector masses:
($M_1$, $M_2$, $\mu$) = (400, 500, 400)~GeV.}
\end{figure}

We display in Figs.~\ref{alphas-075-110-112-100-300-200} and
\ref{alphas-077-110-112-400-500-400} how the allowed regions of the $\alpha$
parameter space are distributed for different choices of the mass parameters and
different cuts on $\tan\beta$.  These plots are obtained by collecting all
points for which solutions (in terms of $\tan\beta, M_{H^\pm}, \alpha_1,
\alpha_2, \alpha_3$) are found during a particular scan (indicated by the
parameters at the top), and binning them in the $\alpha_i$ variables.
In these figures,
we also indicate (yellow plane and green lines) limits where there is no CP
violation.  These figures show that in general, it is much easier to
accommodate CP violation at low values of $\tan\beta$ than at higher values.

%%%%%%%%%%%%%%%%%%%%%%%%%%%%%%%%%%%%%%%%%%%%%%%%%%%%%%%%%%%%%%%%%%%%%%%%%%
%%%%%%%%%%%%%%%%%%%%%%%%%%%%%%%%%%%%%%%%%%%%%%%%%%%%%%%%%%%%%%%%%%%%%%%%%%

%%%%%%%%%%%%%%%%%%%%%%%%%%%%%%%%%%%%%%%%%%%%%%%%%%%%%%%%%%%%%%%%%%%%%%%%%%
%%%%%%%%%%%%%%%%%%%%%%%%%%%%%%%%%%%%%%%%%%%%%%%%%%%%%%%%%%%%%%%%%%%%%%%%%%

In addition to the CP-conserving limits mentioned above there exist also 
other regions in the parameter space which imply CP invariance.
In the following we will identify those regions in terms of the scalar masses, 
$\tan\beta$ and $\alpha_i$.
In this process we will make repeated use of the 
following identities that follow from the orthogonality of $R$:
\begin{eqnarray}
{\cal M}^2_{11}&=&R_{11}^2(M_1^2-M_2^2)+R_{31}^2(M_3^2-M_2^2)+M_2^2\nonumber\\
{\cal M}^2_{22}&=&R_{12}^2(M_1^2-M_2^2)+R_{32}^2(M_3^2-M_2^2)+M_2^2\nonumber\\
{\cal M}^2_{33}&=&R_{13}^2(M_1^2-M_2^2)+R_{33}^2(M_3^2-M_2^2)+M_2^2\nonumber\\
{\cal M}^2_{12}&=&R_{11}R_{12}(M_1^2-M_2^2)+R_{31}R_{32}(M_3^2-M_2^2)
\nonumber\\
{\cal M}^2_{13}&=&R_{11}R_{13}(M_1^2-M_2^2)+R_{31}R_{33}(M_3^2-M_2^2)
\nonumber\\
{\cal M}^2_{23}&=&R_{12}R_{13}(M_1^2-M_2^2)+R_{32}R_{33}(M_3^2-M_2^2)\nonumber
\end{eqnarray}
In addition we order the masses so that $M_1\leq M_2\leq M_3$.
%%%%%%%%%%%%%%%%%%%%%%%%%%%%%%%%%%%%%%%%%%%%%%%%%%%%%%%%%%%%%%%%%%%%%%%%%%%%%
\subsection{$\Im \lambda_5=0$}
%%%%%%%%%%%%%%%%%%%%%%%%%%%%%%%%%%%%%%%%%%%%%%%%%%%%%%%%%%%%%%%%%%%%%%%%%%%%%

Whenever $\Im\lambda_5=0$ we know that CP is conserved. To see when this
happens, we note that
\begin{eqnarray}
-\frac{v^2}{2}s_\beta\Im\lambda_5={\cal M}^2_{13}
= R_{11}R_{13}(M_1^2-M_2^2)+R_{31}R_{33}(M_3^2-M_2^2)
\nonumber\\
-\frac{v^2}{2}c_\beta\Im\lambda_5={\cal M}^2_{23}
= R_{12}R_{13}(M_1^2-M_2^2)+R_{32}R_{33}(M_3^2-M_2^2).
\end{eqnarray}
Both ${\cal M}^2_{13}$ and ${\cal M}^2_{23}$ must vanish in order to get
$\Im\lambda_5=0$. (If only one of these quantities vanishes while the other is
nonzero, that would correspond to $\beta=0$ or $\beta=\pi/2$.)  That happens
in the following cases
\begin{itemize}
\item Full mass degeneracy $M_1=M_2=M_3$.
\item Lower mass degeneracy $M_1=M_2<M_3$, when one of the following 
conditions is satisfied
\begin{itemize}
\item $\alpha_2=\pi/2$ or $\alpha_3=\pi/2$ (This corresponds to $R_{33}=0$.)
\item $\alpha_2=0$ and $\alpha_3=0$. 
(This corresponds to $R_{31}=0$ and $R_{32}=0$.)
\end{itemize}
\item Higher mass degeneracy $M_1<M_2=M_3$, when one of the following 
conditions is satisfied
\begin{itemize}
\item $\alpha_2=0$ (This corresponds to $R_{13}=0$.)
\item $\alpha_2=\pi/2$ (This corresponds to $R_{11}=0$ and $R_{12}=0$.)
\end{itemize}
\item Nondegenerate masses $M_1<M_2<M_3$, when one of the following 
conditions is satisfied
\begin{itemize}
\item $\alpha_2=\pi/2$ 
(This corresponds to $R_{11}=0$ and $R_{12}=0$ and $R_{33}=0$.)
\item $\alpha_2=0$ and $\alpha_3=0$ 
(This corresponds to $R_{13}=0$ and $R_{31}=0$ and $R_{32}=0$.)
\item $\alpha_2=0$ and $\alpha_3=\pi/2$ 
(This corresponds to $R_{13}=0$ and $R_{33}=0$.)
\end{itemize}
\end{itemize}
%%%%%%%%%%%%%%%%%%%%%%%%%%%%%%%%%%%%%%%%%%%%%%%%%%%%%%%%%%%%%%%%%%%%%%%%%%%%%
\subsection{Partially degenerate masses}
%%%%%%%%%%%%%%%%%%%%%%%%%%%%%%%%%%%%%%%%%%%%%%%%%%%%%%%%%%%%%%%%%%%%%%%%%%%%%
We have seen in the previous section that full mass degeneracy implies
$\Im\lambda_5=0$. However, when $\Im\lambda_5\neq 0$, partial mass degeneracy
is still possible in some cases. We will point out those cases here. We begin
by noting that
\begin{eqnarray}
{\cal M}^2_{13}-\tan\beta{\cal M}^2_{23}=0\nonumber
\end{eqnarray}
or, equivalently
\begin{eqnarray}
R_{13}(R_{11}-R_{12}\tan\beta)(M_1^2-M_2^2)
+R_{33}(R_{31}-R_{32}\tan\beta)(M_3^2-M_2^2)=0.\label{eqx}
\end{eqnarray}
\begin{itemize}
\item Lower mass degeneracy $M_1=M_2<M_3$ is allowed when
$R_{31}-R_{32}\tan\beta=0$. (Excluding $R_{31}=R_{32}=0$ which would
imply $\Im \lambda_5=0$ as explained in the previous sub-section.)
\item Higher mass degeneracy $M_1<M_2=M_3$ is allowed when
$R_{11}-R_{12}\tan\beta=0$. (Excluding $R_{11}=R_{12}=0$ which would
imply $\Im \lambda_5=0$ as explained in the previous sub-section.)
\end{itemize}
%%%%%%%%%%%%%%%%%%%%%%%%%%%%%%%%%%%%%%%%%%%%%%%%%%%%%%%%%%%%%%%%%%%%%%%%%%%%%
\subsection{$\lambda_1=\lambda_2$ and $\tan\beta=1$}
%%%%%%%%%%%%%%%%%%%%%%%%%%%%%%%%%%%%%%%%%%%%%%%%%%%%%%%%%%%%%%%%%%%%%%%%%%%%%
We know that even when $\Im\lambda_5\neq 0$ we can have cases of CP
conservation. One such case is when $\lambda_1=\lambda_2$ and $\tan\beta=1$.
\begin{eqnarray}
\left[\lambda_1-\lambda_2\right]_{\tan\beta=1}
&=&\frac{2}{v^2}\left({\cal M}^2_{11}-{\cal M}^2_{22}\right)\nonumber\\
&=&\frac{2}{v^2}
\left[(R_{11}^2-R_{12}^2)(M_1^2-M_2^2)
+(R_{31}^2-R_{32}^2)(M_3^2-M_2^2)\right]\label{eq1}
\end{eqnarray}
This expression must be zero subject to the constraint (\ref{eqx})
in order to have CP conservation.  Thus, when $\Im\lambda_5\neq 0$ and
$\tan\beta=1$ we have CP conservation in the following cases
\begin{itemize}
\item Lower mass degeneracy $M_1=M_2<M_3$ when $R_{31}=R_{32}\neq 0$.
\item Higher mass degeneracy $M_1<M_2=M_3$ when $R_{11}=R_{12}\neq 0$.
\item Nondegenerate masses $M_1<M_2<M_3$ in one of the following cases:
\begin{itemize}
\item When $R_{11}=R_{12}$ and $R_{31}=R_{32}$. 
(Excluding the cases which would imply $\Im\lambda_5=0$.)
This corresponds to $\alpha_1=\pi/4$, $\alpha_3=0$ and $\alpha_2$ arbitrary
(but not 0 or $\pi/2$).
\item When $R_{11}=R_{12}$ and $R_{33}=0$ (this implies $R_{31}=-R_{32}$). 
(Excluding the cases which would imply $\Im\lambda_5=0$.)
This corresponds to $\alpha_1=\pi/4$, $\alpha_3=\pi/2$ and $\alpha_2$ arbitrary
(but not 0 or $\pi/2$).
\item When $R_{31}=R_{32}$ and $R_{13}=0$ (this implies $R_{11}=-R_{12}$). 
(Excluding the cases which would imply $\Im\lambda_5=0$.)
This corresponds to $\alpha_1=-\pi/4$, $\alpha_2=0$ and $\alpha_3$ arbitrary
(but not 0 or $\pi/2$).
\end{itemize}
\end{itemize}
%%%%%%%%%%%%%%%%%%%%%%%%%%%%%%%%%%%%%%%%%%%%%%%%%%%%%%%%%%%%%%%%%%%%%%%%%%%%%
\subsection{$\lambda_1=\lambda_2$ and $(\lambda_1-\lambda_3-\lambda_4)^2
=|\lambda_5|^2$}
%%%%%%%%%%%%%%%%%%%%%%%%%%%%%%%%%%%%%%%%%%%%%%%%%%%%%%%%%%%%%%%%%%%%%%%%%%%%%

Finally, when $\Im\lambda_5\neq 0$ we also have CP conservation when 
$\lambda_1=\lambda_2$ and $(\lambda_1-\lambda_3-\lambda_4)^2=|\lambda_5|^2$.
In order to see what this corresponds to, we start by solving the equation
$\lambda_1=\lambda_2$ for the parameter $\nu$. We find
\begin{eqnarray}
\nu=\frac{s_\beta^2{\cal M}^2_{11}-c_\beta^2{\cal M}^2_{22}}
{v^2(s_\beta^2-c_\beta^2)}
\end{eqnarray}
when $\lambda_1=\lambda_2$.
Furthermore, we find that 
\begin{eqnarray}
(\lambda_1-\lambda_3-\lambda_4)^2-|\lambda_5|^2&=&
\left[\frac{1}{c_\beta^2}\left(\frac{{\cal M}^2_{11}}{v^2}-\nu s_\beta^2\right)
-\frac{{\cal M}^2_{12}}{v^2s_\beta c_\beta}
-\frac{{\cal M}^2_{33}}{v^2}\right]^2\nonumber\\
&&-\left(\frac{{\cal M}^2_{33}}{v^2}-\nu\right)^2
-\frac{4}{v^4}\left[({\cal M}^2_{13})^2+({\cal M}^2_{23})^2\right].
\end{eqnarray}
By substituting the expression for $\nu$ which is valid when 
$\lambda_1=\lambda_2$, we arrive at
\begin{eqnarray}
&&\left[(\lambda_1-\lambda_3-\lambda_4)^2
-|\lambda_5|^2\right]_{\lambda_1=\lambda_2}\nonumber\\
&&=\frac{1}{v^4t_\beta^2(t_\beta^2-1)}\left\{
t_\beta^2(1+t_\beta^2)({\cal M}^2_{22}-{\cal M}^2_{11})
({\cal M}^2_{22}+{\cal M}^2_{11}-2{\cal M}^2_{33})\right.\nonumber\\
&&\hspace*{2cm}
+2t_\beta^3(1+t_\beta^2)({\cal M}^2_{33}-{\cal M}^2_{22}){\cal M}^2_{12}
-2t_\beta(1+t_\beta^2)({\cal M}^2_{33}-{\cal M}^2_{11}){\cal M}^2_{12}
\nonumber\\
&&\hspace*{2cm}
\left.+(t_\beta^2-1)(1+t_\beta^2)^2({\cal M}^2_{12})^2
-4t_\beta^2(t_\beta^2-1)\left[({\cal M}^2_{13})^2+({\cal M}^2_{23})^2\right]
\right\}\nonumber
\end{eqnarray}
After substituting the mass matrix elements into this expression we get:
\begin{eqnarray}
&&\left[(\lambda_1-\lambda_3-\lambda_4)^2
-|\lambda_5|^2\right]_{\lambda_1=\lambda_2}\nonumber\\
&&=\frac{(M_3^2-M_2^2)^2(R_{31}-t_\beta R_{32})}{v^4t_\beta^2(1-t_\beta^2)}
\left\{
(1+t_\beta^2)(R_{31}+t_\beta R_{32})(R_{32}-t_\beta R_{31})^2\right.\nonumber\\
&&\hspace*{5cm}
\left.+2t_\beta R_{33}^2
\left[(R_{31}(t_\beta^3-3t_\beta)+R_{32}(1-3t_\beta^2)\right]\right\}
\nonumber\\
&&+\frac{(M_1^2-M_2^2)^2(R_{11}-t_\beta R_{12})}{v^4t_\beta^2(1-t_\beta^2)}
\left\{
(1+t_\beta^2)(R_{11}+t_\beta R_{12})(R_{12}-t_\beta R_{11})^2\right.\nonumber\\
&&\hspace*{5cm}
\left.+2t_\beta R_{13}^2\left[(R_{11}(t_\beta^3-3t_\beta)
+R_{12}(1-3t_\beta^2)\right]\right\}\nonumber\\
&&+\frac{2(M_1^2-M_2^2)(M_3^2-M_2^2)}{v^4t_\beta^2(1-t_\beta^2)}\left\{
(1+t_\beta^2)(R_{12}-t_\beta R_{11})(R_{32}-t_\beta R_{31})
(R_{11}R_{31}-t_\beta^2R_{12}R_{32})\right.\nonumber\\
&&\hspace*{5cm}+t_\beta(1+t_\beta^2)R_{13}^2
\left[t_\beta(R_{32}^2-R_{31}^2)+R_{31}R_{32}(1-t_\beta^2)\right]\nonumber\\
&&\hspace*{5cm}+t_\beta(1+t_\beta^2)R_{33}^2
\left[t_\beta(R_{12}^2-R_{11}^2)+R_{11}R_{12}(1-t_\beta^2)\right]\nonumber\\
&&\hspace*{5cm}\left.-4t_\beta^2(1-t_\beta^2)R_{13}R_{33}
(R_{11}R_{31}+R_{12}R_{32})\right\}\label{monsterequation}
\end{eqnarray}
This expression must be zero subject to the constraint (\ref{eqx})
in order to have CP conservation.  Thus, when $\Im\lambda_5\neq 0$ and
$\tan\beta\neq 1$ we have CP conservation in the following cases
\begin{itemize}
\item Lower mass degeneracy $M_1=M_2<M_3$ when $R_{31}=\tan\beta R_{32}\neq 0$
and $\mu^2=R_{33}^2M_2^2+(R_{31}^2+R_{32}^2)M_3^2$.
\item Higher mass degeneracy $M_1<M_2=M_3$ when $R_{11}=\tan\beta R_{12}\neq
0$ and $\mu^2=R_{13}^2M_2^2+(R_{11}^2+R_{12}^2)M_1^2$.
\item The case of nondegenerate masses $M_1<M_2<M_3$ yields several 
possibilities for CP conservation. To see this we
rewrite (\ref{eqx}) as
\begin{eqnarray}
\frac{M_3^2-M_2^2}{M_2^2-M_1^2}=\frac{R_{13}(R_{11}-\tan\beta R_{12})}
{R_{33}(R_{31}-\tan\beta R_{32})}
\end{eqnarray}
which must be a positive quantity in the case of nondegenerate
masses. Solving this equation for $M_3^2-M_2^2$ and substituting into
(\ref{monsterequation}) we end up with
\begin{eqnarray}
&&\frac{(M_2^2-M_1^2)^2(R_{31}+t_\beta R_{32})}
{v^4R_{33}^2(R_{31}-t_\beta R_{32})}\,
\frac{(1+t_\beta^2)}{t_\beta^2(1-t_\beta^2)}\,
(R_{11}-t_\beta R_{12})(R_{11}+t_\beta R_{12})
\nonumber\\
&&\times(R_{21}-t_\beta R_{22})(R_{21}+t_\beta R_{22})=0.
\end{eqnarray}
The solutions of this equation that yield CP-conservation are those that
also imply positive values of $(M_3^2-M_2^2)/(M_2^2-M_1^2)$). They are
\begin{itemize}
\item[1.] $R_{11}+t_\beta R_{12}=0$
\item[2.] $R_{21}+t_\beta R_{22}=0$
\item[3.] $R_{31}+t_\beta R_{32}=0$
\end{itemize}
provided
\begin{eqnarray}
\frac{R_{13}(R_{11}-\tan\beta R_{12})}{R_{33}(R_{31}-\tan\beta R_{32})}>0.
\end{eqnarray}

\end{itemize}

%%%%%%%%%%%%%%%%%%%%%%%%%%%%%%%%%%%%%%%%%%%%%%%%%%%%%%%%%%%%%%%%%%%%%%%%%%%%%
\subsection{CP conservation and mass degeneracy}
%%%%%%%%%%%%%%%%%%%%%%%%%%%%%%%%%%%%%%%%%%%%%%%%%%%%%%%%%%%%%%%%%%%%%%%%%%%%%

Summarizing the results presented in this Appendix in terms of mass 
degeneracy we see the following:

\begin{itemize}
\item
When all three masses are degenerate, CP is conserved because
$\Im\lambda_5=0$. For this to happen the two
input masses must be equal, $M_1=M_2$. In most cases (when
$R_{33}(R_{31}-R_{32}\tan\beta)\neq0$) this will lead to $M_3$ being equal to
$M_1$ and $M_2$.  If $R_{33}(R_{31}-R_{32}\tan\beta)=0$ one cannot determine
$M_3$ from the input parameters, but it can be arbitrarily chosen equal to the
two other masses. 

\item
When there is only lower mass degeneracy, $M_1=M_2$ and
$R_{33}(R_{31}-R_{32}\tan\beta)=0$ one cannot determine $M_3$ from the input
parameters, but it can then be arbitrarily chosen to be higher than the two
other masses; $M_1=M_2<M_3$. CP is then conserved if either:
\begin{itemize}
\item $R_{33}=0$ or $R_{31}=R_{32}=0$,
\item $R_{31}=R_{32}\neq0$ and $\tan\beta=1$,
\item $R_{31}=R_{32}\tan\beta\neq0$ and $\tan\beta\neq1$ and 
$\mu^2=R_{33}^2M_2^2+(R_{31}^2+R_{32}^2)M_3^2$.
\end{itemize}
\item
When there is only higher mass degeneracy, CP is conserved in some special
cases. For this to happen we must first choose $M_1<M_2$ and
$R_{13}(R_{11}-R_{12}\tan\beta)=0$. Then, if
$R_{33}(R_{31}-R_{32}\tan\beta)\neq0$, $M_2=M_3$. If
$R_{33}(R_{31}-R_{32}\tan\beta)=0$, one cannot determine $M_3$ from the input
parameters, but it can be arbitrarily chosen to equal $M_2$; $M_1<M_2=M_3$. CP
is then conserved if either:
\begin{itemize}
\item $R_{13}=0$ or $R_{11}=R_{12}=0$,
\item $R_{11}=R_{12}\neq0$ and $\tan\beta=1$,
\item $R_{11}=R_{12}\tan\beta\neq0$ and $\tan\beta\neq1$ and 
$\mu^2=R_{13}^2M_2^2+(R_{11}^2+R_{12}^2)M_1^2$.
\end{itemize}
\item
There are also cases of CP conservation in the mass non-degenerate case
$M_1<M_2<M_3$. For this to happen we must first choose $M_1<M_2$ and 
$\alpha_1$, $\alpha_2$, $\alpha_3$ and $\tan\beta$ in such a way that 
$M_3>M_2$ or in a way such that
$M_3$ cannot be determined from the input parameters. Then $M_3$ can be
arbitrarily chosen to be higher than $M_2$. In both cases $M_1<M_2<M_3$. CP is
then conserved if either:
\begin{itemize}
\item $R_{11}=R_{12}=R_{33}=0$,
\item $R_{31}=R_{32}=R_{13}=0$,
\item $R_{13}=R_{33}=0$,
\item $R_{11}=R_{12}$ and $R_{31}=R_{32}$ 
(excluding the three cases already mentioned above) and $\tan\beta=1$,
\item $R_{11}=R_{12}\neq0$ and $R_{33}=0$ and $\tan\beta=1$,
\item $R_{31}=R_{32}\neq0$ and $R_{13}=0$ and $\tan\beta=1$,
\item $R_{11}+R_{12}\tan\beta=0$ or $R_{21}+R_{22}\tan\beta=0$ or 
$R_{31}+R_{32}\tan\beta=0$ (excluding the three first cases mentioned above) 
and $\tan\beta\neq1$ and $\mu^2$ takes on special values.
\end{itemize}
\end{itemize}

%%%%%%%%%%%%%%%%%%%%%%%%%%%%%%%%%%%%%%%%%%%%%%%%%%%%%%%%%%%%%%%%%%%%%%%%%%%%%
\section{Different basis}
\setcounter{equation}{0}
\setcounter{subsection}{0}
%\renewcommand{\thesection}{B}
%%%%%%%%%%%%%%%%%%%%%%%%%%%%%%%%%%%%%%%%%%%%%%%%%%%%%%%%%%%%%%%%%%%%%%%%%%%%%

For the purpose of determining the electroweak parameters $T$ and $S$ in
Sec.~\ref{Sec:exp-constraints}, we need to relate the rotation matrix $R$
defined by (2.3) in \cite{El Kaffas:2006nt} and the $O$ defined by (59) in
\cite{Grimus:2007if}.  First, we find the $U$ and $V$ (in order to emphasize
that the inert doublet is not yet included, we here adopt a subscript
``2HDM'') of \cite{Grimus:2007if} in the basis adopted in \cite{El
Kaffas:2006nt}, where the doublets are denoted by $\Phi_i$ (see
Eq.~(\ref{Obasis})).

For $\phi_i$ defined in the basis in which only $\phi_1$ has a non-zero
v.e.v.\ (the ``Higgs basis") we have
\begin{equation}
\phi_1=\left(\begin{array}{c}G^+\\ (v+H+iG^0)/\sqrt{2}\end{array}\right) \;\;
\phi_2=\left(\begin{array}{c}H^+\\ (R+iI)/\sqrt{2}\end{array}\right) \;\;
\phi_3=\left(\begin{array}{c}\eta^+\\ (S+iA)/\sqrt{2}\end{array}\right).
\label{Hbasis}
\end{equation}
The following transformation relates $\Phi_i$ and $\phi_j$:
\begin{equation}
\left(\begin{array}{c}\phi_1\\\phi_2\end{array}\right)=
\left(\begin{array}{cc}\cb & \sb \\-\sb & \cb \end{array} \right)
\left(\begin{array}{c}\Phi_1\\\Phi_2\end{array}\right),
\label{Umatdef}
\end{equation}
with $\phi_3=\eta$.
Since in (\ref{Hbasis}) the charged Higgs bosons are mass eigenstates 
(according to (15) in \cite{Grimus:2007if}) we obtain
\begin{equation}
U_\text{2HDM}=\left(\begin{array}{cc}\cb & \sb \\-\sb & \cb 
\end{array} \right)^T.
\label{Umat}
\end{equation}
The matrix $V_\text{2HDM}$ is defined through (see Eq.~(60) in
\cite{Grimus:2007if})
\begin{equation}
\left(\begin{array}{c}\eta_1+i\chi_1\\\eta_2+i\chi_2\end{array} \right)
=V_\text{2HDM}
\left(\begin{array}{c}G^0\\H_1\\H_2\\H_3\end{array}\right)
\label{Vmatdef}
\end{equation}
where $H_i$ are mass eigenstates. The rotation matrix $R$ is defined by
$H_i=R_{ij}\eta_j$ (see Eq.~(\ref{Eq:R-def})). Invoking the relation
\begin{equation}
\left(\begin{array}{c}\chi_1\\
\chi_2\end{array}\right)=\left(\begin{array}{cc}\cb & -\sb \\
\sb & \cb \end{array} \right)
\left(\begin{array}{c}G^0\\\eta_3\end{array}\right)
\end{equation}
and replacing in the L.H.S.\ of (\ref{Vmatdef}) $\eta_{1,2}$ and $\chi_{1,2}$
by the mass eigenstates $G^0$ and $H_j$, then $V_\text{2HDM}$ is identified
as:
\begin{equation}
V_\text{2HDM}=
\left(
\begin{array}{cccc}
i\cb & R_{11}-i\sb R_{13}& R_{21}-i\sb R_{23} & R_{31}-i\sb R_{33}\\
i\sb & R_{12}+i\cb R_{13}& R_{22}+i\cb R_{23} & R_{32}+i\cb R_{33}
\end{array} 
\right).
\label{Vmat}
\end{equation}

%%%%%%%%%%%%%%%%%%%%%%%%%%%%%%%%%%%%%%%%%%%%%%%%%%%%%%%%%%%%%%%%%%%%%%%%%%%%%%%

\end{document}